\documentclass{article}
\usepackage{graphicx}
\usepackage[centertags]{amsmath}
\usepackage{amsfonts}
\usepackage{amssymb}
\usepackage{amsthm}

\title{Generalized Weyl-Wigner-Moyal Formalism\\ and Topological Groups}

\author{J. J. S\l awianowski, V. Kovalchuk, A. Martens,\\ 
B. Go\l ubowska and E. E. Ro\.zko\\
Institute of Fundamental Technological Research,\\
Polish Academy of Sciences,\\
$5^{\rm B}$, Pawi\'{n}skiego str., 02-106 Warsaw, Poland\\
e-mails: jslawian@ippt.gov.pl, vkoval@ippt.gov.pl,\\ 
amartens@ippt.gov.pl, bgolub@ippt.gov.pl, erozko@ippt.gov.pl}

\begin{document}

\maketitle
\begin{abstract}
Discussed are some geometric aspects 
of the phase space formalism in 
quantum mechanics in the sense 
of Weyl, Wigner, Moyal and Ville. We 
analyze the relationship between 
this formalism and geometry of 
the Galilei group, classical momentum 
mapping, theory of unitary projective 
representations of groups, and 
theory of groups algebras. Later 
on we present some generalization 
to quantum mechanics on 
locally compact Abelian groups. It 
is based on Pontryagin duality.
Indicated are certain physical aspects
in quantum dynamics of 
crystal lattices, including the 
phenomenon of ``Umklapp-Prozessen''
\end{abstract}

Let us consider a quantum-mechanical system of classical analogy, with $n$ degrees of freedom. To avoid beginning with a crowd of symbols and abstraction, we postpone some more geometry for later on, and begin with purely analytical formulas. So, the configuration space and the dual space of conjugate momenta are simply identified with $\mathbb{R}^{n}$, and the classical phase space --- with $\mathbb{R}^{2n}$. The standard scalar product in $\mathbb{R}^{n}$ will be denoted by the symbol $\overline{u}\cdot \overline{v}$, and the squared magnitude of $\mathbb{R}^{n}$-vectors --- by $u^{2}=\overline{u}^{2}=\overline{u}\cdot \overline{u}$. The symplectic form $\Gamma$ on $\mathbb{R}^{2n}$ has the matrix
\begin{equation}\label{eq.1}
\Gamma= \left[\begin{array}{cc}
0 & -I \\
I & 0
\end{array}\right],
\end{equation}
so, its evaluation on the pair of $\mathbb{R}^{2n}$-vectors 
\begin{equation}
\overline{z}_{1}=\left(\overline{x}_{1}, \overline{y}_{1}\right), \qquad \overline{z}_{2}=\left(\overline{x}_{2}, \overline{y}_{2}\right),
\qquad \overline{x}_{i}, \overline{y}_{i}\in \mathbb{R}^{n},
\end{equation}
is given by
\begin{equation}\label{eq.2}
\Gamma(\overline{z}_{1}, \overline{z}_{2})=\Gamma((\overline{x}_{1}, \overline{y}_{1}), (\overline{x}_{2}, \overline{y}_{2})):=\overline{y}_{1}\cdot \overline{x}_{2}-\overline{y}_{2}\cdot \overline{x}_{1}.
\end{equation}
Fourier analysis is meant in the convention:
\begin{eqnarray}
\psi(\overline{x})&=&\frac{1}{(2\pi)^{n}}\int \varphi(\overline{k})\exp \left(i\overline{k}\cdot \overline{x}\right)d_{n}\overline{k},\label{eq.3a}\\ 
\varphi(\overline{k})&=&\int \psi(\overline{x})\exp (-i\overline{k}\cdot \overline{x})d_{n}\overline{x},\label{eq.3b}
\end{eqnarray}
where $\overline{k}$ is the wave vector and $\psi$, $\varphi$ are Fourier-related. Often in physics $\varphi$ is referred to 
as the Fourier-profile of $\psi$ and denoted by $\widehat{\psi}$, 
\begin{eqnarray}
\psi(\overline{x})&=&\frac{1}{(2\pi)^{n}}\int \widehat{\psi}(\overline{k})\exp (i\overline{k}\cdot \overline{x})d_{n}\overline{k},\label{eq.4a}\\ 
\widehat{\psi}(\overline{k})&=&\int \psi(\overline{x})\exp (-i\overline{k}\cdot \overline{x})d_{n}\overline{x}.\label{eq.4b}
\end{eqnarray}
This notation stresses the fact that to the extent known from the theory of Fourier transforms, $\psi$ and $\widehat{\psi}$ determine uniquely each other. Roughly speaking, they describe the same "thing", "state" in two different "representations". This is an infinite-dimensional analogue of expressing the same vector in terms of its components with respect to two different bases. The linear formula (\ref{eq.4a}), (\ref{eq.4b}) is just the transformation rule between two representations. 

Quantum linear momentum $\overline{p}$ is proportional to the wave vector $\overline{k}$ and the proportionality coefficient equals the Planck constant $\hbar$,
\begin{equation}\label{eq.5}
\overline{p}=\hbar \overline{k}.
\end{equation}
This is just where this universal constant enters the treatment. Therefore, we write that
\begin{eqnarray}
\psi(\overline{x})&=&\frac{1}{(2\pi\hbar)^{n}}\int \widehat{\psi}[\overline{p}]\exp \left(\frac{i}{\hbar}\overline{p}\cdot \overline{x}\right)d_{n}\overline{p},\label{eq.6a}\\
\widehat{\psi}[\overline{p}]&=&\varphi[\overline{p}]=\int \psi(\overline{x})\exp \left(-\frac{i}{\hbar}\overline{p}\cdot \overline{x}\right)d_{n}\overline{x},\label{eq.6b}
\end{eqnarray}
where, obviously,
\begin{equation}\label{eq.7}
\widehat{\psi}[\hbar \overline{k}]=\widehat{\psi}(\overline{k}).
\end{equation}
The formulas (\ref{eq.6a}), (\ref{eq.6b}), (\ref{eq.7}) belong to the mathematical purity, rather exaggerated from the physicists point of view. Nevertheless, as functions of the indicated arguments, $\widehat{\psi}[\cdot]$, $\widehat{\psi}(\cdot)$ are different. When we use different kinds of brackets, we have symbols of two distinct functions, and everything is logically clean. Of course, when using the formulas carefully, one can avoid this baroque redundancy of decoration. Physicists usually proceed in this way.

As we know from the elementary introduction to quanta, energy and frequency are related to each other like $\overline{p}$ and $\overline{k}$ in (\ref{eq.5}),
\begin{equation}\label{eq.8}
E=\hbar \omega.
\end{equation}
For the classical free particle we have
\begin{equation}\label{eq.9}
E=\frac{\overline{p}^{2}}{2m},
\end{equation}
where $m$ denotes, obviously, the particle mass.

Therefore, (\ref{eq.6a}), (\ref{eq.6b}) is extended to the time evolution in Galilean space-time as follows:
\begin{equation}\label{eq.10}
\psi(t,\overline{x})=\frac{1}{(2\pi\hbar)^{n}}\int \widehat{\psi}[\overline{p}]\exp \left(-\frac{i}{\hbar}\left(\frac{\overline{p}^{2}}{2m}t-\overline{p}\cdot \overline{x}\right)\right)d_{n}\overline{p},
\end{equation}
where the former symbol $\varphi[\overline{p}]$ for the wave profile would be perhaps less confusing,
\begin{equation}\label{eq.11}
\psi(t,\overline{x})=\frac{1}{(2\pi\hbar)^{n}}\int \varphi[\overline{p}]\exp \left(-\frac{i}{\hbar}\frac{\overline{p}^{2}}{2m}t\right)\exp \left(\frac{i}{\hbar}\overline{p}\cdot \overline{x}\right)d_{n}\overline{p}.
\end{equation}

This was just one of heuristic, de Broglie-like way of "deriving" the free Schr\"{o}dinger equation, because it is clear that (\ref{eq.10}), (\ref{eq.11}) satisfies
\begin{equation}\label{eq.12}
i\hbar\frac{\partial \psi}{\partial t}=-\frac{\hbar^{2}}{2m}\Delta \psi,
\end{equation}
where, obviously, $\Delta$ denotes the $n$-dimensional Laplace operator,
\begin{equation}\label{eq.13}
\Delta=\delta^{ij}\frac{\partial^{2}}{\partial x^{i}\partial x^{j}}=\sum_{i}\frac{\partial^{2}}{\partial \left(x^{i}\right)^{2}}.
\end{equation}

Introducing the momentum operator, i.e., quantum linear momentum 
\begin{equation}\label{eq.14}
{\bf P}_{a}=\frac{\hbar}{i}\frac{\partial}{\partial x^{a}},
\end{equation}
we can write (\ref{eq.12}) as follows:
\begin{equation}\label{eq.15}
i\hbar\frac{\partial \psi}{\partial t}={\bf T} \psi, \qquad {\bf T}=\frac{1}{2m}\delta^{ab}\mathbf{P}_{a}\mathbf{P}_{b}.
\end{equation}

When some potential $V$ is admitted, (\ref{eq.15}) is replaced by
\begin{equation}\label{eq.16}
i\hbar\frac{\partial \psi}{\partial t}={\bf H} \psi={\bf T}\psi+V\psi,
\end{equation}
where ${\bf H}$ and ${\bf T}$ are respectively the total energy (Hamiltonian) and kinetic energy operators.

In configuration space $\mathbb{R}^{n}$ the group of translations, just the additive group $\mathbb{R}^{n}$ itself acts through regular translations:
\begin{equation}\label{eq.17}
\overline{\alpha}\in \mathbb{R}^{n}: \qquad \mathbb{R}^{n}\ni \overline{x}\mapsto \overline{x}+\overline{\alpha}\in \mathbb{R}^{n}.
\end{equation}
Obviously, this action is transitive, free and effective. The same group of spatial translations acts in the phase space $\mathbb{R}^{2n}\simeq \mathbb{R}^{n}\times \mathbb{R}^{n}$,
\begin{equation}\label{eq.18}
\overline{\alpha}\in \mathbb{R}^{n}: \qquad \mathbb{R}^{2n}\ni (\overline{x}, \overline{p})\mapsto (\overline{x}+\overline{\alpha},\overline{p});
\end{equation}
obviously, this action is not transitive any longer, but
it is still free and effective.

In non-relativistic quantum mechanics, time plays the parameter role. The group $\mathbb{R}^{n}$ of Galilean boosts is realized in the velocity space $\mathbb{R}^{n}$ as follows:
\begin{equation}
\overline{\nu} \in \mathbb{R}^{n}: \qquad \mathbb{R}^{n} \ni \overline{v} \mapsto \overline{v} + \overline{\nu} \in \mathbb{R}^{n}. \label{eq.19}
\end{equation}
This action is represented in the space of positions and velocities, i.e., in the space of Newtonian states $P_{N}$ and in the phase space $P$ of positions and canonical momenta, on this level both analytically identified with $\mathbb{R}^{2n}$, as follows:
\begin{eqnarray}
\overline{\nu} \in \mathbb{R}^{n}:&\quad& \mathbb{R}^{2n} \ni \left(\overline{x}, \overline{v}\right) \mapsto \left(\overline{x}, \overline{v} + \overline{\nu}\right) \in \mathbb{R}^{2n},\label{eq.20a} \\
\overline{\nu} \in \mathbb{R}^{n}:&\quad& \mathbb{R}^{2n} \ni \left(\overline{x}, \overline{p}\right) \mapsto \left(\overline{x}, \overline{p} + m\overline{\nu}\right) \in \mathbb{R}^{2n}. \label{eq.20b}
\end{eqnarray}
These actions are also effective and free, but evidently non-transitive. The coefficient $m$ in (\ref{eq.20b}) follows evidently from the Legendre transformation for the free material point. 

Seemingly, one might parameterize (\ref{eq.20b}) by the boost linear momenta $\overline{\pi}$: 
\begin{equation}
\overline{\pi} \in \mathbb{R}^{n}: \qquad \mathbb{R}^{2n} \ni \left(\overline{x}, \overline{p}\right) \mapsto \left(\overline{x},\overline{p} + \overline{\pi}\right) \in \mathbb{R}^{2n}. \label{eq.21}
\end{equation}
Physically, the boost transformations are parameterized by the velocity increase $\overline{\nu}$, and analytically they are expressed in terms of the Galilean space-time coordinates $(t, \overline{x})$ as follows:
\begin{equation}
\overline{\nu} \in \mathbb{R}^{n}: \qquad \mathbb{R}^{n+1} \ni \left(t, \overline{x}\right) \mapsto \left(t, \overline{x} + \overline{\nu}t\right) \in \mathbb{R}^{n+1}. \label{eq.22}
\end{equation}

Spatial translations and boosts, i.e., (\ref{eq.17}), (\ref{eq.20a}), (\ref{eq.20b}), generate the $2n$-dimen\-sional Abelian group of phase-space translations:
\begin{equation}
\left(\overline{\alpha}, \overline{\nu}\right) \in \mathbb{R}^{2n}: \qquad \mathbb{R}^{2n} \ni \left( \overline{x}, \overline{p}\right) \mapsto \left(\overline{x} + \overline{\alpha}, \overline{p} + m\overline{\nu}\right) \in \mathbb{R}^{2n}. \label{eq.23}
\end{equation}
In Galilean space-time terms we have that
\begin{equation}
\mathbb{R}^{n+1} \ni \left(t, \overline{x}\right) \mapsto \left(t, \overline{x} + \overline{\alpha}+ \overline{\nu}t\right) \in \mathbb{R}^{n+1}. \label{eq.24}
\end{equation}

It is an important subgroup of the total Galilei group of the dimension $(n+1)(n+2)/2$; physically $10$, because $n=3$. In a sense, it was a starting point of the reasoning of Weyl, which resulted in what is today referred to as the Weyl-Wigner-Moyal-Ville formalism \cite{Moyal, Weyl, Weyl2, Wigner}. The original paper by H. Weyl was entitled {\it "Quantum kinematics as an Abelian group of rotations"} \cite{Weyl2}. That Abelian group of rotations was just some unitary realization of the above group of classical phase-space translations. On the classical level one is dealing here with the non-exact momentum mapping. Namely, classical Hamiltonian generators of spatial translations are given by the components of canonical momenta $p_{k}$, whereas the boost transformations are generated by $b_{k}=m x_{k}$. The group of space-time translations (\ref{eq.23}) is evidently Abelian and its generating vector fields form an Abelian Lie algebra under the Lie brackets operation. But, obviously, the Hamiltonian generators $p_{k}$, $b_{k}$ do not form Poisson-involutive system. It is clear, we have here $2n$ functionally independent functions, whereas it is $n$ that is the maximal possible number of Jacobi-independent functions in involution. It is just the mass $m$ that is an obstacle against the existence of the exact momentum mapping for (\ref{eq.23}),
\begin{equation}
\left\{b_{k}, p_{l}\right\} = \left\{m x_{k}, p_{l}\right\} = m \delta_{kl}. \label{eq.25}
\end{equation}
There is a deep convolution of concepts here: Galilei group, foundations of quanta, phase-space aspects of quantum mechanics and its group-algebraic formulation.

Let us now just go back to quantum mechanics and review the quantum counterparts of the above transformation groups.

Spatial translations act on wave functions in the position representation as follows, through the translation operators $U(\overline{\alpha})$:
\begin{equation}
\left(U(\overline{\alpha})\psi\right)(\overline{x}):= \psi \left(\overline{x} - \overline{\alpha} \right). \label{eq.26}
\end{equation}
Galilean boosts are naturally defined in a similar way as translations $V(\overline{\nu})$ of wave profiles in the momentum representation:
\begin{equation}
\left(V(\overline{\nu})\varphi \right)(\overline{p}):= \varphi \left(\overline{p} - m \overline{\nu} \right), \label{eq.27}
\end{equation}
or, using the suggestive symbols $\widehat{\psi}$:
\begin{equation}
\left(V(\overline{\nu})\widehat{\psi} \right)[\overline{p}]:= \widehat{\psi} \left[\overline{p} - m \overline{\nu} \right]. \label{eq.28}
\end{equation}

Using the "dictionary" (\ref{eq.6a}), (\ref{eq.6b}) we obtain immediately the following coordinate representation:
\begin{eqnarray}
\left(V(\overline{\nu})\psi \right)(\overline{x})&=& \exp \left(\frac{i}{\hbar} m \overline{\nu} \cdot \overline{x}\right)\psi(\overline{x})\nonumber \\
&=& \exp \left(\frac{i}{\hbar} \overline{\pi} \cdot \overline{x}\right)\psi(\overline{x}). \label{eq.29}
\end{eqnarray}
To be very precise, we might distinguish between symbols $V(\overline{\nu})$ in (\ref{eq.27}), (\ref{eq.28}) and (\ref{eq.29}). However, this would lead to an obscure crowd of symbols. By abuse of language, we write as above instead of using, e.g., two different symbols like $V(\overline{\nu})$, $\widetilde{V}(\overline{\nu})$, related to each other through the Fourier transformation:
\begin{equation}
\left(V(\overline{\nu})\psi \right)^{\wedge}= \left(\widetilde{V}(\overline{\nu})\widehat{\psi} \right). \label{eq.30}
\end{equation}
The both realizations (\ref{eq.26}) and (\ref{eq.29}) are unitary in the sense of the standard scalar product of wave functions:
\begin{eqnarray}
\left\langle  \psi_{1}| \psi_{2} \right\rangle  = \int \overline{\psi_{1}}(\overline{x})\psi_{2}(\overline{x})d_{n}\overline{x}&=&  \frac{1}{(2 \pi \hbar)^{n}} \int \overline{\widehat{\psi}_{1}}[\overline{p}]\widehat{\psi}_{2}[\overline{p}]d_{n}
\overline{p} \nonumber\\ 
&=& \frac{1}{(2 \pi)^{n}} \int \overline{\widehat{\psi}_{1}}(\overline{k})\widehat{\psi}_{2}(\overline{k})d_{n}
\overline{k}. \label{eq.31}
\end{eqnarray}
Unitarity is obvious:
\begin{eqnarray}
\left\langle  U(\overline{\alpha})\psi_{1} \left|\right. U(\overline{\alpha})\psi_{2} \right\rangle  = \left\langle  \psi_{1}| \psi_{2} \right\rangle , \label{eq.32}\\
\left\langle  V\left(\overline{\nu}\right)\psi_{1}|V\left(\overline{\nu}\right)\psi_{2}\right\rangle  =\left\langle  \psi_{1}|\psi_{2}\right\rangle . \label{eq.33}
\end{eqnarray}
The rule (\ref{eq.32}) is a direct consequence of the invariance of the
Lebesgue measure on $\mathbb{R}^{n}$ under additive translations
$\overline{x}\mapsto\overline{x}+\overline{\alpha}$. One deals
here just with the special case of Haar measure.

Obviously, we have the following representation properties:
\begin{eqnarray}
U(\overline{\alpha}_{1})U(\overline{\alpha}_{2}) & = & U(\overline{\alpha}_{1}+\overline{\alpha}_{2}),\label{eq.34a}\\
V(\overline{\nu}_{1})V(\overline{\nu}_{2}) & = & V(\overline{\nu}_{1}+\overline{\nu}_{2}).\label{eq.34b}
\end{eqnarray}

Some problems appears on the level of the quantum counterpart of (\ref{eq.23}). And this is very heart of quantum mechanics: non-commutativity
of certain operations, uncertainty relations, etc. On the classical
level this problem is reflected by the failure of the momentum mapping.
On the quantum level this is the problem of nontrivial projective
unitary representations, non-reducible to the usual vector representations. In any case we construct the operations $U(\overline{\alpha})V(\overline{\nu})$
and $V(\overline{\nu})U(\overline{\alpha})$,
\begin{eqnarray}
\left(U(\overline{\alpha})V(\overline{\nu})\psi\right)\left(\overline{x}\right)
&=&\exp\left(\frac{i}{\hbar}m\overline{\nu}\cdot\left(\overline{x}-
\overline{\alpha}\right)\right)\psi\left(\overline{x}-\overline{\alpha}\right),
\label{eq.35a}\\
\left(V(\overline{\nu})U(\overline{\alpha})\psi\right)\left(\overline{x}\right)
&=&\exp\left(\frac{i}{\hbar}m\overline{\nu}\cdot\overline{x}\right)\psi
\left(\overline{x}-\overline{\alpha}\right).\label{eq.35b}
\end{eqnarray}

Let us now define 
\begin{equation}
\mathcal{W}\left(\overline{\alpha},\overline{\nu}\right):=
U(\overline{\alpha})V(\overline{\nu})=\exp\left(-\frac{i}{\hbar}
m\overline{\nu}\cdot\overline{x}\right)V(\overline{\nu})
U(\overline{\alpha}).\label{eq.36}
\end{equation}
After some calculations one can show that 
\begin{eqnarray}
\mathcal{W}\left(\overline{\alpha}_{1},\overline{\nu}_{1}\right)
\mathcal{W}\left(\overline{\alpha}_{2},\overline{\nu}_{2}\right)&=&
\exp\left(\frac{i}{\hbar}m\overline{\nu}_{1}\cdot\overline{\alpha}_{2}
\right)\mathcal{W}(\overline{\alpha}_{1}+\overline{\alpha}_{2},
\overline{\nu}_{1}+\overline{\nu}_{2}),\label{eq.37}\\
\mathcal{W}\left(\overline{\alpha},\overline{\nu}\right)^{-1}&=&
\exp\left(\frac{i}{\hbar}m\overline{\nu}\cdot\overline{\alpha}
\right)\mathcal{W}(-\overline{\alpha},-\overline{\nu}).\label{eq.38}
\end{eqnarray}
Using the non-physical parametrization in terms of the boost momenta
$\overline{\pi}=m\overline{\nu}$, we would write:
\begin{equation}
\mathcal{W}\left[\overline{\alpha}_{1},\overline{\pi}_{1}\right]
\mathcal{W}\left[\overline{\alpha}_{2},\overline{\pi}_{2}\right]=
\exp\left(\frac{i}{\hbar}m\overline{\pi}_{1}\cdot\overline{\alpha}_{2}\right)
\mathcal{W}\left[\overline{\alpha}_{1}+\overline{\alpha}_{2},
\overline{\pi}_{1}+\overline{\pi}_{2}\right].\label{eq.39}
\end{equation}
It is seen from (\ref{eq.37}) that assignment 
\begin{equation}
\mathbb{R}^{2n}\ni(\overline{\alpha},\overline{\nu})\mapsto
\mathcal{W}(\overline{\alpha},\overline{\nu})
\end{equation}
is not a unitary representation in the literal sense, nevertheless,
it is a projective unitary representation, i.e., representation up
to the unimodular factor. The formulas (\ref{eq.37}), (\ref{eq.38}) are non-aesthetic. Modifying (\ref{eq.36}) by an appropriate unimodular factor, we transform them into more natural canonical form, where the exponent in the projective factor is skew-symmetric in its arguments, and the inverse operation
is obtained by inverting the sign of the argument \cite{Bargmann}.
Namely, let us define:
\begin{equation}
\mathbb{W}\left(\overline{\alpha},\overline{\nu}\right)=
\exp\left(\frac{i}{\hbar}mg\overline{\nu}\cdot\overline{\alpha}\right)
\mathcal{W}(\overline{\alpha},\overline{\nu}),\label{eq.40}
\end{equation}
where $g$ is some coefficient. We try to choose it in such a way
as to obtain the elegant rule:
\begin{equation}
\mathbb{W}\left(\overline{\alpha},\overline{\nu}\right)^{-1}=
\mathbb{W}\left(-\overline{\alpha},-\overline{\nu}\right).\label{eq.41}
\end{equation}
After easy calculations we obtain that
\begin{equation}
g=\frac{1}{2},\label{eq.42}
\end{equation}
and then
\begin{equation}
\mathbb{W}\left(\overline{\alpha},\overline{\nu}\right)=
\exp\left(\frac{i}{2\hbar}m\overline{\nu}\cdot\overline{\alpha}\right)
\mathcal{W}(\overline{\alpha},\overline{\nu}).\label{eq.43}
\end{equation}
In addition we have the following nice composition rule:
\begin{equation}
\mathbb{W}\left(\overline{\alpha}_{1},\overline{\nu}_{1}\right)
\mathbb{W}\left(\overline{\alpha}_{2},\overline{\nu}_{2}\right)=
\exp\left(\frac{i}{2\hbar}m\left(\overline{\nu}_{1}\cdot\overline{\alpha}_{2}
-\overline{\nu}_{2}\cdot\overline{\alpha}_{1}\right)\right)\mathbb{W}
\left(\overline{\alpha}_{1}+\overline{\alpha}_{2},\overline{\nu}_{1}+
\overline{\nu}_{2}\right),\label{eq.44}
\end{equation}
or briefly,
\begin{equation}
\mathbb{W}\left(\overline{u}_{1}\right)\mathbb{W}\left(\overline{u}_{2}\right)=
\exp\left(\frac{i}{2\hbar}m\Gamma\left(\overline{u}_{1},\overline{u}_{2}\right)
\right)\mathbb{W}(\overline{u}_{1}+\overline{u}_{2}),\label{eq.45}
\end{equation}
where $\Gamma$ is the standard symplectic bilinear form on $\mathbb{R}^{2n}$,
analytically based on the skew-symmetric matrix (\ref{eq.1}) denoted by the
same symbol, 
\begin{equation}\label{eq.46}
\Gamma\left(\overline{u}_{1},\overline{u}_{2}\right)=u_{1}^{T}\Gamma u_{2}=\Gamma_{ab}u_{1}^{a}u_{2}^{b}=\overline{\nu}_{1}\overline{\alpha}_{2}
-\overline{\nu}_{2}\overline{\alpha}_{1},
\end{equation} 
and $\overline{u}_{i}=\left(\overline{\alpha}_{i},\overline{\nu}_{i}\right)$, $i=1,2$. Using the $\overline{\pi}$-parametrization we would have that
\begin{eqnarray}
\mathbb{W}\left[\overline{z}_{1}\right]\mathbb{W}\left[\overline{z}_{2}\right] 
&=&\mathbb{W}\left[\overline{\alpha}_{1},\overline{\pi}_{1}\right]
\mathbb{W}\left[\overline{\alpha}_{2},\overline{\pi}_{2}\right]\nonumber\\
&=&\exp\left(\frac{i}{2\hbar}\left(\overline{\pi}_{1}\overline{\alpha}_{2}-
\overline{\pi}_{2}\overline{\alpha}_{1}\right)\right)
\mathbb{W}\left[\overline{\alpha}_{1}
+\overline{\alpha}_{2},\overline{\pi}_{1}+\overline{\pi}_{2}\right]\nonumber \\
&=&\exp\left(\frac{i}{2\hbar}\Gamma\left(\overline{z}_{1},\overline{z}_{2}
\right)\right)\mathbb{W}\left[\overline{z}_{1}+\overline{z}_{2}\right].
\label{eq.47} 
\end{eqnarray}
In this way, according to the V. Bargmann procedure \cite{Bargmann}, the symmetric
part of the projective factor exponent is eliminated as a kind of
non-physical gauge.

The Weyl-Wigner-Moyal-Ville formalism is based on the objects $\mathbb{W}(\overline{z})$, i.e., $\left(\overline{\alpha},\overline{\pi}_{}\right)$-parametrization, on the phase-space language. But, as mentioned, if we use wave functions as defined on the Galilei space, the parametrization of boosts based on velocities $\overline{\nu}$ is just the proper language. It is well known that in the homogeneous formulation of classical mechanics or, as J. L. Synge calls it, in the $QT$-space formulation \cite{Synge1, Synge2}, energy in a sense plays the role of the negative canonical momentum conjugate to the time variable,
\begin{equation}\label{eq.48}
p_{t}=-E.
\end{equation}
This is readably expressed by the Hamilton-Jacobi equation,
\begin{equation}\label{eq.49}
\frac{\partial S}{\partial t}+H\left(t,q^{i};\frac{\partial S}{\partial q^{i}}\right)=0,
\end{equation}
because the physical meaning of Hamiltonian $H$ is just the energy.

However, in Galilean space-time, unlike in the relativistic Minkowski space, the system $\left(-E,\overline{p}\right)$ is not a four-dimensional covariant vector. It is not a linear object at all, however, it is a geometric object of the affine group. It transforms under Galilean boosts according to affine, i.e., linear-inhomogeneous rule:
\begin{eqnarray}
&&\left[E,\overline{p}\right]\mapsto\left[E',\overline{p}'\right]=
\left['E,'\overline{p}\right]+\left[\frac{m}{2}\overline{\nu}^{2},
m\overline{\nu}\right],\label{eq.50}\\
&&\left['E,'\overline{p}\right]=\left[E+\overline{\nu}\cdot\overline{p},
\overline{p}\right].\label{eq.51}
\end{eqnarray}
The second expression (\ref{eq.51}) is just the homogeneous part or four-covector part of the transformation rule. The second term on the right-hand side of (\ref{eq.50}) is the inhomogeneous correction term. Let us notice its interesting structure: it is the kinetic energy and linear momentum assigned by the mass $m$ to the boost velocity parameter.

The corresponding rule for the harmonic plane wave
\begin{equation}\label{eq.52}
\psi_{\left(E,\overline{p}\right)}=\exp\left(-\frac{i}{\hbar}\left(
Et-\overline{p}\cdot\overline{x}\right)\right)
\end{equation}
is
\begin{equation}\label{eq.53}
\psi_{\left(E,\overline{p}\right)}\mapsto\psi_{\left(E',\overline{p}'
\right)}=:\psi',
\end{equation}
where $\left(E,\overline{p}\right)$, $\left(E',\overline{p}'\right)$ are related to each other by (\ref{eq.50}), (\ref{eq.51}),
\begin{equation}
\psi'\left(t,\overline{x}\right)=\exp\left(-\frac{i}{\hbar}m\left(
\frac{\overline{\nu}^{2}}{2}t-\overline{\nu}\cdot\overline{x}\right)\right)
\psi\left(t,\overline{x}-\overline{\nu}t\right).
\end{equation}
Then, making use of (\ref{eq.10}), we find the general rule for boosts in terms of the Galilei space-time wave functions:
\begin{equation}\label{eq.54}
\left(V\{\overline{\nu}\}\psi\right)\left(t,\overline{x}\right):=
\exp\left(-\frac{i}{\hbar}m\left(\frac{\overline{\nu}^{2}}{2}t-
\overline{\nu}\cdot\overline{x}\right)\right)\psi\left(t,\overline{x}-
\overline{\nu}t\right).
\end{equation}
Translations act in the usual way:
\begin{equation}\label{eq.55}
\left(U\{\overline{\alpha}\}\psi\right)\left(t,\overline{x}\right)=
\psi\left(t,\overline{x}-\overline{\alpha}\right).
\end{equation}
Unlike (\ref{eq.55}), the formula (\ref{eq.54}) unifies the argument-wise action with the unimodular factor. This is evidently related to the fact that it is not usual (vector) unitary representations but unitary projective representations that is physically meaningful for the Galilei group as a symmetry of nonrelativistic quantum mechanics. There is an essential difference between the status of mass in non-relativistic and relativistic quantum theories. In relativistic theory, mass is, by some abuse of language, an eigenvalue of one of Casimir invariants of the Poincare group, namely the Lorentzian square of the four-momentum. In non-relativistic theory, mass is a label of the projective unitary representation of the Galilei group. This is one of infinity of qualitative discontinuities of the correspondence and limit transition between relativistic and nonrelativistic theories. Classically this problem is represented by the non-exactness of the momentum mapping. Namely, the $2n$-dimensional group of space-time translations in $\mathbb{R}^{2n}$ is an Abelian subgroup of the Galilei group. Nevertheless, its Hamiltonian generators are not in Poisson-involution, and this failure of involution is measured, labelled by the particle's mass $m$. This fact has obvious counterparts on the quantum level. For "small" values of the parameters $\overline{\alpha}$, $\overline{\nu}$ we have the obvious expansions:
\begin{eqnarray}
U\left(\overline{\alpha}\right)&\approx&{\rm Id}-\alpha^{k}\frac{\partial}{\partial x^{k}}={\rm Id}-\alpha^{k}\frac{i}{\hbar}P_{k},\qquad P_{k}=\frac{\hbar}{i}\frac{\partial}{\partial x^{k}},\label{eq.56}\\
V\left(\overline{\nu}\right)&\approx&{\rm Id}+\nu^{k}\frac{i}{\hbar}mx^{k}={\rm Id}+\nu^{k}\frac{i}{\hbar}B_{k}={\rm Id}+\nu^{k}\frac{i}{\hbar}mQ_{k},\label{eq.57}
\end{eqnarray}
where $Q_{k}$ and $P_{k}$ are respectively the position and momentum operators. They satisfy the well-known quantum Poisson brackets:
\begin{equation}\label{eq.58}
\frac{1}{i\hbar}\left[Q_{a},P_{b}\right]=\delta_{ab}.
\end{equation}
Obviously, ${\rm Id}$ in (\ref{eq.56}), (\ref{eq.57}) denotes the identity operator. For finite values of parameters we may write
\begin{eqnarray}
U(\overline{\alpha})&=&\exp\left(-\frac{i}{\hbar}\alpha^{k}P_{k}\right),
\label{eq.59a}\\
V(\overline{\nu})&=&\exp\left(\frac{i}{\hbar}\nu^{k}B_{k}\right)=
\exp\left(\frac{i}{\hbar}m\nu^{k}Q_{k}\right),\label{eq.59b}
\end{eqnarray}
with all obvious provisos concerning domains of unbounded operators and the convergence of $\exp$-series. Unlike this, (\ref{eq.26}) and (\ref{eq.27}) are always well defined, bounded and unitary \cite{Woron}. Mathematically the step (\ref{eq.59a}), (\ref{eq.59b}) is artificial and superfluous, nevertheless, one uses it for physical reasons, because $Q_{k}$, $P_{k}$ are physically interpreted as operators (unbounded, however) of physical quantities (positions and linear momenta).

The adjoint rule action in the algebra of operators is given by
\begin{eqnarray}
{\rm Ad}_{U(\overline{\alpha})}\varrho:=U(\overline{\alpha})\varrho
U(\overline{\alpha})^{-1},&\quad& {\rm ad}_{P_{k}}\varrho=\frac{1}{i\hbar}\left[P_{k},\varrho\right],\label{eq.61}\\
{\rm Ad}_{V(\overline{\nu})}\varrho:=V(\overline{\nu})\varrho
V(\overline{\nu})^{-1},&\quad& {\rm ad}_{Q_{k}}\varrho=\frac{1}{i\hbar}\left[Q_{k},\varrho\right],\label{eq.62}
\end{eqnarray}
and more generally,
\begin{eqnarray}
{\rm Ad}_{\mathbb{W}(\overline{u})}\varrho&=&\mathbb{W}(\overline{u})\varrho
\mathbb{W}(\overline{u})^{-1},\label{eq.63}\\
{\rm Ad}_{\mathbb{W}[\overline{z}]}\varrho&=&\mathbb{W}[\overline{z}]\varrho
\mathbb{W}[\overline{z}]^{-1}.\label{eq.64}
\end{eqnarray}
It is obvious that transformations (\ref{eq.63}), (\ref{eq.64}) form a $(2n)$-dimensional Abelian group of the $\mathbb{R}^{2n}$-topology. The projective factor which appears when acting on wave functions is cancelled there. The operators $\varrho$ in (\ref{eq.61}), (\ref{eq.62}), (\ref{eq.63}), (\ref{eq.64}) may be physically interpreted, e.g., as von Neumann density operators (density matrices) or physical quantities. Therefore, the above formulas are transformation rules for quantum states and observables. The position and momentum operators do not commute with each other, but similarity transformations generated by them, i.e., translations and Galilean boosts do commute. This is just what H. Weyl had in mind when using the formulation {\it "Quantum kinematics as the Abelian group of rotations"} (obviously "rotations" in the sense "unitary transformations", or rather, "unitary similarities") \cite{Weyl, Weyl2}. Let us stress that it is not $\mathbb{W}(\overline{u})/\mathbb{W}(z)$-transformations that are commutative, but rather the corresponding similarities:
\begin{equation}
\left[{\rm Ad}_{\mathbb{W}(\overline{u}_{1})}, {\rm Ad}_{\mathbb{W}(u_{2})}\right]=0, \qquad 
\left[{\rm Ad}_{\mathbb{W}(\overline{z}_{1})}, {\rm Ad}_{\mathbb{W}(\overline{z}_{2})}\right]=0. \label{eq.65}
\end{equation}
Similarly, $Q_{k}$ and $P_{k}$ do not commute with each other, but ${\rm ad}_{Q_{k}}$ and ${\rm ad}_{P_{k}}$ are commutative:
\begin{equation}
\left[{\rm ad}_{Q_{k}}, {\rm ad}_{P_{i}}\right]=0. \label{eq.66}
\end{equation}
This is some interesting and quite non-accidental convolution of problems: Gali\-lean symmetry, momentum mapping, quantum mechanics in the phase space, projective unitary transformations. 

Let us stress that the commutativity on the level of von Neumann operators (\ref{eq.65}), (\ref{eq.66}) is lifted to the floor of wave functions so as to result in the following commutation rule, the finite version of canonical commutation rules \cite{Bargmann, Woron}:
\begin{equation}
\mathbb{W}[z_{1}] \mathbb{W}[z_{2}] \mathbb{W}[z_{1}]^{-1} \mathbb{W}[z_{2}]^{-1}= \exp\left( \frac{i}{\hbar}\Gamma(z_{1}, z_{2})\right)
{\rm Id}, \label{eq.67}
\end{equation}
or, equivalently, with the explicit use of the mass factor in the exponent:
\begin{equation}
\mathbb{W}(u_{1}) \mathbb{W}(u_{2}) \mathbb{W}(u_{1})^{-1} \mathbb{W}(u_{2})^{-1}= \exp\left( \frac{i}{\hbar}m \Gamma(u_{1}, u_{2})\right){\rm Id}. \label{eq.68}
\end{equation}

Let us notice that here there are no problems with domains, because one deals with bounded operators, without the artificial introducing of non-bounded differential operators and the resulting complicated discussion of the problem on which function space the rules like (\ref{eq.58}) are well defined. Obviously, (\ref{eq.58}) is an infinitesimal version of the finite rule (\ref{eq.67}). In the latter rule one uses the group commutator of the type $g_{1}g_{2}g_{1}^{-1}g_{2}^{-1}$, where $g_{1}$, $g_{2}$ are group elements.

It is instructive to take the linear shell of the family of operators $\mathbb{W}(u)$ or $\mathbb{W}[z]$. This is a rather general procedure. When one deals with the representation $G \ni g \mapsto U(g) \in L(H)$ of some locally compact topological group $G$ in a linear space $H$ (in quantum mechanics one deals mainly with unitary representations), then it is natural to consider the algebra of operators:
\begin{equation}
A:= \int \alpha(g)U(g)dg, \label{eq.69}
\end{equation}
where the integration symbol $dg$ refers to the Haar measure on $G$ and $\alpha$ is an integrable function, i.e., an element of $L^{1}(G)$. When $G$ is a Lie group, one admits also distributions on $G$. Important physical quantities are described by operators (\ref{eq.69}). Multiplying operator (\ref{eq.69}) by another one of this shape,
\begin{equation}
B:= \int \beta(g)U(g)dg, \label{eq.70}
\end{equation}
we obtain
\begin{equation}
AB= \int \left(\alpha \ast \beta \right)(g)U(g)dg, \label{eq.71}
\end{equation}
where $\alpha \ast  \beta$ is the group convolution of functions,
\begin{equation}
\left(\alpha \ast  \beta \right)(g)= \int \alpha(h) \beta\left(h^{-1} g\right) dh. \label{eq.72}
\end{equation}
In this way we study the problem in terms of the group algebra over $G$ \cite{Woron}. The same may be repeated for the projective unitary representation. Multiplication of operators is then represented by so-called "twisted convolution". When this is done for the above assignment $z \mapsto \mathbb{W}[z]$, one obtains in principle the Weyl-Wigner-Moyal-Ville prescription.

So, taking the "continuous linear shell", i.e., superposing continuously all operators $\mathbb{W}[\overline{z}] = \mathbb{W}[\overline{\alpha}, 
\overline{\pi}]$, we obtain the following family of operators:
\begin{eqnarray}
\mathbf{A} &=& \int \widehat{A}(\overline{\alpha}, \overline{\pi}) \mathbb{W}[\overline{\alpha}, \overline{\pi}] d_{n}\overline{\alpha} \frac{d_{n}\overline{\pi}}{(2 \pi \hbar)^{n}} \nonumber\\
&=& \int \widehat{A} (\overline{\alpha}, \overline{\pi}) \exp \left(\frac{i}{\hbar} \left( \pi_{a}\mathbf{Q}^{a} + \alpha^{a}\mathbf{P}_{a} \right)\right)d_{n} \overline{\alpha}\frac{d_{n}\overline{\pi}}{(2 \pi \hbar)^{n}}, \label{eq.73}
\end{eqnarray}
where $\mathbf{Q}^{a}$, $\mathbf{P}_{a}$ denote the above position and momentum operators. The first formula is in principle more correct because it uses only bounded, unitary operators. However, the second version is more close to the classical theory and better suited to the quantum-classical correspondence. Namely, the corresponding expression for the classical phase-space function $A$ through its Fourier transform $\widehat{A}$ (defined, by duality, on the same phase-space), has the form:
\begin{equation}
A\left(\overline{Q}, \overline{P}\right)= \int \widehat{A} (\overline{\alpha}, \overline{\pi}) \exp \left(\frac{i}{\hbar} \left( \pi_{a}Q^{a} + \alpha^{a}P_{a} \right)\right)d_{n} \overline{\alpha}\frac{d_{n}\overline{\pi}}{(2 \pi \hbar)^{n}}. \label{eq.74}
\end{equation}
The formulas (\ref{eq.73}), (\ref{eq.74}) establish the correspondence between classical phase-space functions $A$ and operators $\mathbf{A}$. This is just the Weyl-Wigner-Moyal-Ville representation. The associative products of operators, and their quantum Poisson brackets,
\begin{equation}
\mathbf{A}\mathbf{B},\qquad 
\left\{\mathbf{A},\mathbf{B}\right\}_{\rm quant}= \frac{1}{i\hbar}\left[\mathbf{A}, \mathbf{B}\right]= 
\frac{1}{i\hbar}\left(\mathbf{A}\mathbf{B} - \mathbf{B}\mathbf{A}\right), \label{eq.75}
\end{equation}
become then some expressions in terms of phase space functions, namely, the so-called "star-product",
\begin{equation}
(A\star B)(z)=2^{2n}\int \exp \left(\frac{2i}{\hbar}\Gamma(z-z_{1}, z-z_{2}) \right)A(z_{1})B(z_{2})d\mu(z_{1})d\mu(z_{2}),\label{eq.76a}
\end{equation}
and Moyal-Weyl-Wigner-Ville bracket, 
\begin{equation}
\{A, B\}_{\rm quant}=\frac{1}{i\hbar}(A\star B -B \star A),\label{eq.76b}
\end{equation}
where $\mu$ is the phase-space measure normalized as follows:
\begin{equation}\label{eq.77}
d\mu(z)=d\mu(q, p)=\frac{1}{(2\pi \hbar)^{n}}dq^{1} \ldots dq^{n}dp_{1} \ldots dp_{n}.
\end{equation}
The star-product (\ref{eq.76a}) is not to be confused with the convolution symbol $\ast$ on $\mathbb{R}^{2n}$. But the resulting composition rule for the Fourier transforms $\widehat{A}$, $\widehat{B}$ is somehow related to this convolution, namely, it is a "twisted" convolution in the sense of being modified by the factors of projective representations of the Galilei group. The product (\ref{eq.76a}) is non-local in the phase-space $\mathbb{R}^{2n}$. Obviously, this has to do with the uncertainly relations \cite{JJS-phase,JJS-Heis}.

Let us mention the relationship with the usual operator language. Operators 
${\bf A}$ may be expressed as integral operators with kernels $A[\overline{q}, \overline{q}']$, i.e., 
\begin{equation}\label{eq.78}
({\bf A}\psi)(\overline{q})=\int A[\overline{q}, \overline{q}']\psi(q')d_{n}q'.
\end{equation}
Obviously, very often this representation is rather symbolic in the sense that kernels are distributions, not functions. It is so for the identity operator ${\bf I}$, the kernel of which is given by Dirac-delta,
\begin{equation}\label{eq.79}
I[\overline{q}, \overline{q}']=\delta(\overline{q}- \overline{q}').
\end{equation}
The same holds for unbounded operators of position and differential (thus, also unbounded) operators like linear momentum. Kernels of differential operators are expressed by derivatives of Dirac-delta. The kernels of ${\bf I}$, ${\bf Q}^{i}$, ${\bf P}_{i}$ are respectively given by (\ref{eq.79}) and
\begin{equation}\label{eq.80}
Q^{i}[\overline{q}, \overline{q}']=q^{i}\delta\left(\overline{q}-\overline{q}'
\right),\qquad P_{i}[\overline{q}, \overline{q}']=\frac{\hbar}{i}\frac{\partial}{\partial q^{i}}\delta(\overline{q}-\overline{q}').
\end{equation}

Let us quote the general formulas:
\begin{eqnarray}
A[\overline{q}, \overline{q}']&=&\int \exp \left( \frac{i}{\hbar}\overline{p}\cdot(\overline{q}-\overline{q}')\right)
A\left(\frac{1}{2}(\overline{q}+\overline{q}'), \overline{p} \right)\frac{d_{n}\overline{p}}{(2\pi \hbar)^{n}},\label{eq.81}\\
\label{eq.82}
A(q, p)&=&\int \exp \left( -\frac{i}{\hbar}\overline{p}\cdot\overline{\alpha}\right)
A\left[\overline{q}+\frac{\overline{\alpha}}{2}, \overline{q}-\frac{\overline{\alpha}}{2} \right]d_{n}\overline{\alpha}.
\end{eqnarray}
These relationships are translationally-invariant in the position $\overline{q}$-variables and this is correct, because this means that they are formulated in the configuration space on the basis of affine geometry, not linear ones. Working, from the very beginning, analytically in $\mathbb{R}^{n}$ instead of in an affine space, one obscures this important problems. One might have various artificial and misleading ideas concerning replacing the two-argument configuration functions $A[\overline{q}, \overline{q}']$ by the phase-space functions $A(\overline{q}, \overline{p})$. For example, to perform the Fourier analysis with respect to the first or second argument in $A[\overline{q}, \overline{q}']$. This would be non-aesthetic, translationally non-invariant, and artificial. Probably even just misleading, because of the placing the Planck constant in an improper position. This may damage the quasiclassical analysis. The failure of translational invariance is not only non-aesthetic, but probably also incorrect (it may have to do with entering the Planck constant in an improper way). Configurational translational invariance in (\ref{eq.81}), (\ref{eq.82})) is meant in the sense that the substitution
\begin{equation}\label{eq.83}
A[\overline{q}, \overline{q}']\mapsto A[\overline{q}+\overline{\alpha}, \overline{q}'+\overline{\alpha}]
\end{equation}
is equivalent to the following substitution:
\begin{equation}\label{eq.84}
A(\overline{q}, \overline{p})\mapsto A[\overline{q}+\overline{\alpha}, \overline{p}].
\end{equation}
For the completeness let us write once more the formula (\ref{eq.76a}) explicitly in a purely analytical way:
\begin{eqnarray}\label{eq.85}
(A\star B)\left(q^{i}, p_{i}\right)&=&\left(\frac{2}{2\pi \hbar}\right)^{2n}
\int \exp\left(\frac{2i}{\hbar}
\left(\left(p"_{i}-p_{i}\right)(q'^{i}-q^{i})\right.\right.\\
&-&\left.\left.\left(p'_{i}-p_{i}\right)\left(q"^{i}-q^{i}\right)\right)\right)
A(q', p')B(q", p")d_{n}q' d_{n}p'd_{n}q"
d_{n}p".\nonumber
\end{eqnarray}
In (\ref{eq.76a}), (\ref{eq.85}) the Planck constant is explicitly inserted where it should be. In functions $A$, $B$ it is absent. Let us also quote another formula, alternative to (\ref{eq.81}), (\ref{eq.82})) and explicitly connecting operators ${\bf A}$ with their phase-space functions $A$,
\begin{equation}\label{eq.86}
({\bf A\psi})(q)=\frac{1}{(2\pi \hbar)^{n}}\int \exp \left(\frac{i}{\hbar}\overline{p}\cdot\left(\overline{q}-\overline{q}'\right)\right)
A\left(\frac{1}{2}(\overline{q}+\overline{q}'),\overline{p}\right)d_{n}
\overline{q}'d_{n}\overline{p}.
\end{equation}
Configurational translational invariance is obvious, because $(\overline{q}+\overline{q}')/2$ is an affine concept, namely, centre of the segment between points $\overline{q},\overline{q}'\in \mathbb{R}^{n}$. The linear concepts of $\mathbb{R}^{n}$ are used only apparently.

The Weyl prescription ${\bf A} \leftrightarrow A$ carries over the algebraic structure of the set of operators ${\bf A}$ to the set of phase-space functions $A$. Obviously, the isomorphism is valid for some subsets only, but we shall not go into such details here. By the very correspondence, the star product is associative and bilinear with respect to its factors, in particular, distributive with respect to addition: 
\begin{eqnarray}
(A\star B)\star C&=&A\star (B\star C),\label{eq.87a}\\
(\lambda A+\mu B)\star C&=&\lambda A\star C+\mu B\star C,\label{eq.87b}\\
A\star (\lambda B+\mu C)&=&\lambda A\star B+ \mu A\star C,\label{eq.87c}
\end{eqnarray}
assuming, of course, that the left- and right-hand sides do exist.

Hermitian conjugation of operators is represented by the complex conjugation of functions:
\begin{equation}\label{eq.88}
({\bf A} \leftrightarrow A)\Leftrightarrow ({\bf A}^{+} \leftrightarrow \overline{A}),
\end{equation}
therefore, Hermitian operators (and symmetric ones, formally Hermitian) are represented by real functions, and the following holds:
\begin{equation}\label{eq.89}
\overline{A \star B}=\overline{B}\star \overline{A}.
\end{equation}
The constant function $1$ satisfies 
\begin{equation}\label{eq.90}
1\star A=A\star 1=A
\end{equation}
for any function $A$, and it is the only function of this property.

For any function $A$ the following holds: 
\begin{equation}
\overline{A}\star  A\neq 0,\label{eq.91}
\end{equation}
unless $A$ vanishes almost everywhere. 

For any functions $A$ and $B$ the following holds: 
\begin{equation}
\int A\star  B d\mu=\int ABd\mu,\label{eq.92}
\end{equation}
although in general $A\star  B$ is different than $AB$. But, attention:
in general it is not true for the triple product in the integrand:
\begin{equation}
\int A\star  B\star  Cd\mu\neq\int ABCd\mu.\label{eq.93}
\end{equation}
The trace of operators, if it exists, may be expressed as follows:
\begin{equation}
{\rm Tr}\;\mathbf{A}=\int A\left(\overline{z}\right)d\mu\left(\overline{z}\right)=\int A\left(\overline{q},\overline{p}\right)d_{n}\overline{q}
\frac{d_{n}\overline{p}}{\left(2\pi\right)^{n}}.\label{eq.94}
\end{equation}
Similarly, the scalar product of operators, if it exist (it does not need generally to exist, this is an infinite dimension), is given by: 
\begin{equation}
\left\langle  \mathbf{A},\mathbf{B}\right\rangle  ={\rm Tr}\left(A^{+}B\right)=\int\overline{A(z)}B(z)d\mu(z)=\left\langle  A,B\right\rangle .\label{eq.95}
\end{equation}
The $H^{+}$-algebraic compatibility between this scalar product and
Hermitian conjugation of operators is expressed as follows in two
equivalent ways: 
\begin{eqnarray}
\left\langle  \mathbf{CA},\mathbf{B}\right\rangle  &=&\left\langle  \mathbf{A},\mathbf{C}^{+}\mathbf{B}\right\rangle  =\left\langle  \mathbf{C},\mathbf{B}\mathbf{A}^{+}\right\rangle ,\label{eq.96}\\
\left\langle  C\star  A,B\right\rangle  &=&\left\langle  A,\overline{C}\star  B\right\rangle  =\left\langle  C,B\star  \overline{A}\right\rangle .
\label{eq.97}
\end{eqnarray}
Being isomorphic with the operator multiplication, the star-product
is non-commutative, there are, however, obvious exceptions, e.g.,
when functions depend on the same kind of variables, i.e., only $q$-variables or only $p$-variables, 
\begin{eqnarray}
A(q)\perp B(q) & = & B(q)A(q) = \left(AB\right)(q),\label{eq.98a}\\
A(qp)\perp B(p) & = & B(p)A(p) = \left(AB\right)(p).\label{eq.98b}
\end{eqnarray}
In this special case it reduces to the usual pointwise product. In
general it is non-commutative and non-local in $\mathbb{R}^{2n}$,
or, in a sense, expressed by differential operators of infinite order,
i.e., pseudo-differential operators. In special, simple situations,
the corresponding series reduce to finite-order expressions, e.g.,
\begin{eqnarray}
q^{a}\star  p_{b} & = & q^{a}p_{b}+\frac{i\hbar}{2}\delta^{a}{}_{b},
\label{eq.99a}\\
q^{a}\star  A & = & q^{a}A+\frac{i\hbar}{2}\frac{\partial A}{\partial p_{a}},\label{eq.99b}\\
p_{a}\star  A & = & p_{a}A-\frac{i\hbar}{2}\frac{\partial A}{\partial q^{a}},\label{eq.99c}
\end{eqnarray}
etc. The star product is invariant under translations in $\mathbb{R}^{2n}$,
i.e., defined essentially in the $2n$-dimensional affine space. It
is also invariant under symplectic transformations acting in
$\mathbb{R}^{2n}$. Let us write this precisely. Take an arbitrary
vector $y\in\mathbb{R}^{2n}$ and arbitrary symplectic transformation
$L\in {\rm Sp}(n,\mathbb{R})\subset {\rm GL}(2n,\mathbb{R})$, 
\begin{equation}
\Gamma_{cd}L^{c}{}_{a}L^{d}{}_{b}=\Gamma_{ab}.\label{eq.100}
\end{equation}
They act on the phase-space functions according to the rule:
\begin{equation}
\left(U(L,y)f\right)(z):=f\left(Lz+y\right).\label{eq.101}
\end{equation}
It is obvious that 
\begin{equation}
\left(U(L,y)f\right)\star  \left(U(L,y)g\right)=U\left(L,y\right)(f\star  g).
\label{eq.102}
\end{equation}
Incidentally, this is an interesting quantum aspect of the classical
symplectic group, or more generally, of affine canonical transformations. 

Non-locality of the star-product is of the phase-space volume order
$\hbar^{n}$. Let us remind that, as Landau and Lifshitz stressed, this
is the order of the phase-space volume per one quantum state \cite{LL, Rubinowicz}. The value of $A\star  B$ in a neighborhood of any point $z\in\mathbb{R}^{2n}$ depends essentially on the behaviour of $A$, $B$ in a neighborhood of $z$ of the volume of the order $\hbar^{n}$.

Quantum statistical states are described by density operators/density
matrices $\boldsymbol{\varrho}$. They are Hermitian, normalized, and
positive in the following sense:
\begin{equation}
\boldsymbol{\varrho}^{+}=\boldsymbol{\varrho},\qquad {\rm Tr}\;\boldsymbol{\varrho}=1,\qquad \left\langle  \boldsymbol{\varrho}|\mathbf{A}^{+}\mathbf{A}\right\rangle  ={\rm Tr}\left(\boldsymbol{\varrho}\mathbf{A}^{+}\mathbf{A}\right)\geq 0
\label{eq.103}
\end{equation}
for any operator $A$. This means that $\boldsymbol{\varrho}$ has a non-negative
spectrum and 
\begin{equation}
{\rm Tr}\;\boldsymbol{\varrho}^{2}\leq {\rm Tr}\;\boldsymbol{\varrho}=1.\label{eq.104}
\end{equation}

The particular situations of the pure state corresponds to the situation
when $\boldsymbol{\varrho}$ is a projection onto some one-dimensional
subspace of wave functions, spanned by some wave functions $\psi$,
so that 
\begin{equation}
\mathbf{\boldsymbol{\varrho}}=|\psi><\psi|,\qquad
\varrho\left[\overline{q},\overline{q}'\right]=
\psi\left(\overline{q}\right)\overline{\psi\left(\overline{q}'\right)}.
\label{eq.105}
\end{equation}

Statistical indeterminancy (diffusion) of the quantum state is given
by entropy functional
\begin{equation}
S\left[\boldsymbol{\varrho}\right]=-{\rm Tr}\left(\boldsymbol{\varrho}
\ln\boldsymbol{\varrho}\right).\label{eq.106}
\end{equation}
It evidently vanishes for pure states. The spectrum of $\boldsymbol{\varrho}$
for pure states contains one unit and zeros. Obviously, this is
the regular Hilbert space situation. In practice one uses also non-normalizable
densities just like non-normalizable wave functions. For pure states
(\ref{eq.105}) the corresponding phase-space function $\varrho$ on
$\mathbb{R}^{2n}$, i.e., the Weyl-Moyal-Wigner-Ville distributions, is given
by
\begin{equation}
\varrho\left(\overline{q},\overline{p}\right)=\frac{1}{\left(2\pi\right)^{n}}
\int\overline{\psi}\left(\overline{q}-\frac{1}{2}\hbar\overline{\tau}\right)
\exp\left(- i \overline{\tau}\cdot\overline{p}\right)\psi\left(\overline{q}+
\frac{1}{2}\hbar\overline{\tau}\right)d_{n}\overline{\tau}.\label{eq.107}
\end{equation}

It is important that, as a rule, $\varrho$ is not positive
in the usual pointwise sense. The same concerns the general, not necessarily
pure-state Weyl-Moyal-Wigner-Ville densities. Instead, they are positive
in the sense of positive eingenvalues, i.e., positive expectation
values on all positive quantities, 
\begin{equation}\label{eq.108}
\left\langle  \varrho|\overline{A}\star  A\right\rangle  =\int\varrho\left(z\right)\left(\overline{A}\star  A\right)\left(z\right)d_{\mu}\left(z\right)>0
\end{equation}
for any phase-space function $A$, except of vanishing one, when $A$
vanishes almost everywhere.

Obviously, for any physical quantity $\mathbf{A}=\mathbf{A}^{+}$,
$A=\overline{A}$ the real quantity
\begin{equation}\label{eq.109}
{\rm Tr}\left(\boldsymbol{\varrho}\mathbf{A}\right)=\int\varrho(z)A(z)d\mu(z)
=\langle  A\rangle _{\varrho}
\end{equation}
is an expectation value on the quantum state described by $\boldsymbol{\varrho}$, $\varrho$. Similarly, for any pair of pure states $\boldsymbol{\varrho}_{1}$, $\boldsymbol{\varrho}_{2}$, or equivalently $\varrho_{1}$, $\varrho_{2}$, the quantity
\begin{equation}\label{eq.110}
P\left(\boldsymbol{\varrho}_{1},\boldsymbol{\varrho}_{2}\right)={\rm Tr}\left(\boldsymbol{\varrho}_{1},\boldsymbol{\varrho}_{2}\right)=
\int\varrho_{1}(\overline{z})\varrho_{2}(\overline{z})d\mu(\overline{z})=
|\langle \psi_{1}|\psi_{2}\rangle |^{2}
\end{equation}
is the probability that the state $\boldsymbol{\varrho}_{1}$ will be detected in the state $\boldsymbol{\varrho}_{2}$ or conversely (transition probability). There exists also some formula for mixed states. It is, however, much more complicated, although the expression (\ref{eq.110}) is a first step towards constructing it. It is interesting that (\ref{eq.109}) is identical with the corresponding formula from classical statistical mechanics, in spite that $\varrho$ is not the usual probability distribution localized in the phase space $\mathbb{R}^{2n}$.

Let us remind that there are special cases of Wigner-Ville distributions which are positive also in the usual pointwise sense. They have to do with the harmonic oscillator and coherent states. They are labelled by space-time parameters $\left(\overline{\xi},\overline{\pi}\right)\in\mathbb{R}^{2n}$ and given by
\begin{eqnarray}
E_{\left(\overline{\xi},\overline{\pi}\right)}\left(\overline{q},\overline{p}
\right):=\frac{1}{\left(\pi\hbar\right)^{n}}\exp\left(
-\frac{1}{\hbar}\left(a_{lm}\left(q^{l}-\xi^{l}\right)\left(q^{m}-\xi^{m}\right)
\right.\right.
\nonumber\\
+\left.\left. a^{lm}\left(p_{l}-\pi_{l}\right)\left(p_{m}-\pi_{m}\right)\right)\right),
\label{eq.111}
\end{eqnarray}
where $a$ is interpreted as a kind of metric and the upper-case indices refer to the contravariant inverse
\begin{equation}\label{eq.112}
a_{lm}a^{mk}=\delta_{l}{}^{k}.
\end{equation}
To avoid too much "sophisticated" symbols we can simply take $a_{lm}=\delta_{lm}$, $a^{mk}=\delta^{mk}$ and write
\begin{equation}\label{eq.113}
E_{\left(\overline{\xi},\overline{\pi}\right)}\left(\overline{q},
\overline{p}\right)=\frac{1}{\left(\pi\hbar\right)^{n}}\exp\left(
-\frac{1}{\hbar}\left(\left(\overline{q}-\overline{\xi}\right)^{2}+
\left(\overline{p}-\overline{\pi}\right)^{2}\right)\right).
\end{equation}
Those are Gauss functions and they evidently describe some normalized pure states,
\begin{eqnarray}
\int E_{\left(\overline{\xi},\overline{\pi}\right)}\left(\overline{q},
\overline{p}\right)d_{n}q\frac{d_{n}p}{\left(2\pi\hbar\right)^{n}}&=&1,\label{eq.114}\\
E_{\left(\overline{\xi},\overline{\pi}\right)}\star  
E_{\left(\overline{\xi},\overline{\pi}\right)}&=&
E_{\left(\overline{\xi},\overline{\pi}\right)}.\label{eq.115}
\end{eqnarray}
For any Wigner function $\varrho$, the quantity
\begin{equation}\label{eq.116}
P\left(\varrho,E_{\overline{\zeta}}\right)=\int \varrho(\overline{z})
E_{\overline{\zeta}}(\overline{z})d\overline{z},\qquad \overline{\zeta}=\left(\overline{\xi},\overline{\pi}\right),
\end{equation}
is the detection probability of $E_{\overline{\zeta}}$. As $E_{\overline{\zeta}}$ is strongly concentrated around $\overline{\zeta}$, this quantity, with $\overline{\zeta}$ as a variable, may be considered as a kind of the density probability for $\overline{\zeta}$. $E_{\overline{\zeta}}$ is essentially concentrated in the $\hbar^{n}$-volume cell around $\overline{\zeta}$. The quantity
\begin{eqnarray}
\widetilde{\varrho}(\overline{z})=\widetilde{\varrho}(\overline{q},\overline{p})
&=&\int E_{\overline{z}}\left(\overline{\zeta}\right)\varrho
\left(\overline{\zeta}\right)d\mu\left(\overline{\zeta}\right)\nonumber\\
&=&\int E_{\overline{q},\overline{p}}\left(\overline{\xi},\overline{\pi}\right)\varrho
\left(\overline{\xi},\overline{\pi}\right)d_{n}\overline{\xi}
\frac{d_{\overline{\pi}}}{\left(2\pi\hbar\right)^{n}}\label{eq.117}
\end{eqnarray}
may be interpreted as a kind of positively-definite (in the usual pointwise sense) probability distribution for finding the system in the position $\overline{z}=\left(\overline{q},\overline{p}\right)$ in the phase space $\mathbb{R}^{2n}$. This is so-called Husimi probability distribution. Its disadvantage is that it has bad marginal properties, namely,
\begin{eqnarray}
\int \widetilde{\varrho}(\overline{q},\overline{p})d_{n}\overline{q}&\neq&
\overline{\widehat{\psi}}\left(\overline{p}\right)\widehat{\psi}\left(
\overline{p}\right)=\left|\widehat{\psi}\left(\overline{p}\right)
\right|^{2},\label{eq.118}\\
\int \widetilde{\varrho}(\overline{q},\overline{p})\frac{d_{n}\overline{p}}
{\left(2\pi\hbar\right)^{n}}&\neq&
\overline{\psi}\left(\overline{q}\right)\psi\left(
\overline{q}\right)=\left|\psi\left(\overline{q}\right)
\right|^{2}.\label{eq.119}
\end{eqnarray}
Unlike this, the Wigner-Ville distribution for pure states has good marginal properties,
\begin{eqnarray}
\int \varrho(\overline{q},\overline{p})d_{n}\overline{q}&=&
\overline{\widehat{\psi}}\left(\overline{p}\right)\widehat{\psi}\left(
\overline{p}\right)=\left|\widehat{\psi}\left(\overline{p}\right)
\right|^{2},\label{eq.120}\\
\int \varrho(\overline{q},\overline{p})\frac{d_{n}\overline{p}}
{\left(2\pi\hbar\right)^{n}}&=&
\overline{\psi}\left(\overline{q}\right)\psi\left(
\overline{q}\right)=\left|\psi\left(\overline{q}\right)
\right|^{2}.\label{eq.121}
\end{eqnarray}
But as a probability distribution in the phase space $\mathbb{R}^{2n}$ it is non-local because it may take on negative values. Positive probabilities may be obtained only after averaging, coarse-graining, e.g., like in (\ref{eq.117}) over the phase-space cells of the volume remarkably larger than $\hbar^{n}$, "phase-space volume" of a single quantum state \cite{LL, Rubinowicz}.

It was mentioned above about $H^{+}$-algebraic structures. Any complete system of states $|i\rangle $ in the Hilbert space of wave functions, gives rise to the natural complete systems of operators
\begin{equation}\label{eq.122}
\boldsymbol{\varrho}_{ij}=|i\rangle \langle  j|
\end{equation}
in the operator algebra over the mentioned Hilbert space of wave functions. Then we have the obvious $H^{+}$-algebraic basic relations:
\begin{eqnarray}
\boldsymbol{\varrho}_{ij}^{+}=\boldsymbol{\varrho}_{ji},&\quad&
\boldsymbol{\varrho}_{ij}\boldsymbol{\varrho}_{kl}=\delta_{jk}
\boldsymbol{\varrho}_{il},\label{eq.123}\\
\overline{\varrho_{ij}}=\varrho_{ji},&\quad&
\varrho_{ij}\star  \varrho_{kl}=\delta_{jk}
\varrho_{il}.\label{eq.124}
\end{eqnarray}
But it is also very interesting to use the continuous families of "states" like $|\overline{p}\rangle $ and $|\overline{q}\rangle $, although they do not belong to the Hilbert space. What is even worse, in the position representation $|\overline{q}\rangle $ are even not functions, just some Dirac distributions. Of course, everything may be rigorously formulated, e.g., in the language of "rigged Hilbert spaces". But it is not our aim here; for us it is sufficient to use some symbolic expressions.

The position and momentum representations give rise to the continua of non-normalizable "states" outside the $L^{2}$-framework, $|\overline{q}\rangle $, $|\overline{p}\rangle $, and then to the continua of phase-space "functions" (distributions, to be more precise) of the form:
\begin{eqnarray}
\varrho_{\overline{q}_{1}\overline{q}_{2}}(\overline{q},\overline{p})&=&
\delta\left(\overline{q}-\frac{1}{2}\left(\overline{q}_{1}+
\overline{q}_{2}\right)\right)\exp\left(\frac{i}{\hbar}\overline{p}
\cdot\left(\overline{q}_{2}-\overline{q}_{1}\right)\right),\label{eq.125a}\\
\varrho_{\overline{p}_{1}\overline{p}_{2}}(\overline{q},\overline{p})&=&
\delta\left(\overline{p}-\frac{1}{2}\left(\overline{p}_{1}+
\overline{p}_{2}\right)\right)\exp\left(\frac{i}{\hbar}
\left(\overline{p}_{1}-\overline{p}_{2}\right)\cdot\overline{q}\right).
\label{eq.125b}
\end{eqnarray}
Obviously, they satisfy the following symbolic relationships:
\begin{eqnarray}
\varrho_{\overline{q}_{1}\overline{q}_{2}}\star  \varrho_{\overline{q}_{3}
\overline{q}_{4}}&=&\delta\left(\overline{q}_{2}-\overline{q}_{3}\right)
\varrho_{\overline{q}_{1}\overline{q}_{4}},\label{eq.126}\\
\varrho_{\overline{p}_{1}\overline{p}_{2}}\star  \varrho_{\overline{p}_{3}
\overline{p}_{4}}&=&\delta\left(\overline{p}_{2}-\overline{p}_{3}\right)
\varrho_{\overline{p}_{1}\overline{p}_{4}},\label{eq.127}\\
\varrho_{\overline{q}_{1}\overline{q}_{2}}=\overline{\varrho_{\overline{q}_{2}
\overline{q}_{1}}},&\quad&\varrho_{\overline{p}_{1}\overline{p}_{2}}=
\overline{\varrho_{\overline{p}_{2}\overline{p}_{1}}}.
\end{eqnarray}
All this is done by abuse of language, just like the "states" $|\overline{q}\rangle $, $|\overline{p}\rangle $ and the corresponding relationships:
\begin{eqnarray}
|\overline{q}_{1}\rangle \langle \overline{q}_{2}|\overline{q}_{3}\rangle 
\langle \overline{q}_{4}|&=&\delta\left(\overline{q}_{2}-\overline{q}_{3}
\right)|\overline{q}_{1}\rangle \langle \overline{q}_{4}|,\label{eq.128}\\
|\overline{p}_{1}\rangle \langle \overline{p}_{2}|\overline{p}_{3}\rangle 
\langle \overline{p}_{4}|&=&\delta\left(\overline{p}_{2}-\overline{p}_{3}
\right)|\overline{p}_{1}\rangle \langle \overline{p}_{4}|,\label{eq.129}\\
\left(|\overline{q}_{1}\rangle \langle \overline{q}_{2}|\right)^{+}=
|\overline{q}_{2}\rangle \langle \overline{q}_{1}|,&\qquad&
\left(|\overline{p}_{1}\rangle \langle \overline{p}_{2}|\right)^{+}=
|\overline{p}_{2}\rangle \langle \overline{p}_{1}|.\label{eq.130}
\end{eqnarray}
In analogy to the discrete expansions
\begin{equation}\label{eq.131}
\boldsymbol{\varrho}=\sum_{ij}\varrho_{ij}|i\rangle \langle  j|,\qquad \varrho_{ij}=\langle  i|\boldsymbol{\varrho}|j\rangle ,
\end{equation}
we have the following continuous "expansions" with the use of kernels of integral operators:
\begin{eqnarray}
\boldsymbol{\varrho}&=&\int\varrho\left[\overline{q},\overline{q}'\right]
|\overline{q}\rangle \langle \overline{q}'|d_{n}\overline{q}
d_{n}\overline{q}',\label{eq.132}\\
\boldsymbol{\varrho}&=&\int\varrho\left[\overline{p},\overline{p}'\right]
|\overline{p}\rangle \langle \overline{p}'|\frac{d_{n}\overline{p}}
{\left(2\pi\hbar\right)^{n}}\frac{d_{n}\overline{p}'}
{\left(2\pi\hbar\right)^{n}},\label{eq.133}
\end{eqnarray}
where, obviously,
\begin{equation}\label{eq.134}
\varrho\left[\overline{q},\overline{q}'\right]=\langle  \overline{q}|\boldsymbol{\varrho}|\overline{q}'\rangle ,\qquad \varrho\left[\overline{p},\overline{p}'\right]=\langle  \overline{p}|\boldsymbol{\varrho}|\overline{p}'\rangle .
\end{equation}
In terms of the basic distributions (\ref{eq.125a}), (\ref{eq.125b}) we have the following "continuous expansions" for the corresponding phase-space functions:
\begin{eqnarray}
\textbf{A}&=&\int A\left[\overline{q},\overline{q}'\right]\varrho_{\overline{q},\overline{q}'} d_{n}\overline{q}d_{n}\overline{q}',\label{eq.135}\\
A&=&\int A\left[\overline{p},\overline{p}'\right]
\varrho_{\overline{p},\overline{p}'}\frac{d_{n}\overline{p}}
{\left(2\pi\hbar\right)^{n}}\frac{d_{n}\overline{p}'}
{\left(2\pi\hbar\right)^{n}}.\label{eq.136}
\end{eqnarray}
They are generally valid for phase-space functions, not necessarily for the Wig\-ner-Weyl-Moyal-Ville "quasiprobability" distributions.

It is clear that "diagonal" elements of the $H^{+}$-algebraic "bases" describe "states" of continuous spectra, localized respectively in positions and momenta:
\begin{eqnarray}
\varrho_{\overline{x},\overline{x}}\left(\overline{q},\overline{p}\right)&=&
\delta\left(\overline{q}-\overline{x}\right),\label{eq.137}\\
\varrho_{\overline{y},\overline{y}}\left(\overline{q},\overline{p}\right)&=&
\delta\left(\overline{p}-\overline{y}\right).\label{eq.138}
\end{eqnarray}
These exceptional Wigner densities for pure states are localized, concentrated on the corresponding $n$-dimensional Lagrange manifolds in the $2n$-dimensional phase space $\mathbb{R}^{2n}$. The relationship between quantum and classical concepts is here very direct and explicitly seen.

It is no longer the case for Wigner-Weyl-Moyal-Ville functions (\ref{eq.107}) for the general pure states.

Nevertheless, the relationship does exist in the quasiclassical limit $\hbar\rightarrow 0$. Namely, if we express the wave function as follows:
\begin{equation}\label{eq.139}
\psi(q)=\sqrt{D(q)}\exp\left(\frac{i}{\hbar}S(q)\right),
\end{equation}
and assume the real functions $D$, $S$ to be $\hbar$-independent (or at least to have well-defined limits when $\hbar\rightarrow 0$), then, in the limit $\hbar\rightarrow 0$, obviously, the limit meant in the distribution sense, one obtains for (\ref{eq.107}):
\begin{equation}\label{eq.140}
\varrho_{\rm cl}\left[D,S\right]=\lim_{\hbar\rightarrow 0}\varrho
\left[D,S\right]=D(\overline{q})\delta\left(p_{1}-\frac{\partial S}{\partial q_{1}}\right)\ldots\delta\left(p_{n}-\frac{\partial S}{\partial q_{n}}\right).
\end{equation}
This is a distribution the support of which is the $n$-dimensional Lagrangian submanifold $\mathfrak{M}_{S}\subset \mathbb{R}^{2n}$ given by equations \cite{JJS-phase}:
\begin{equation}\label{eq.141}
p_{i}-\frac{\partial S}{\partial q_{i}}=0,\qquad i=1,\ldots,n.
\end{equation}
The phase-space point $\left(\overline{q},\overline{p}\right)=\left(\overline{q},\nabla S_
{\overline{q}}\right)\in\mathfrak{M}_{S}$ is taken with the statistical weight $D\left(\overline{q}\right)$ meant as a continuous probability distribution.

Obviously, there are plenty of warnings and provisos concerning these formulas. First of all, everything is correct only when $S$ is well defined, i.e., in points where the absolute value $D$ is non-vanishing. And, of course, in regions remote from the classical turning points, and classically admissible.

It is clear that the "basic distributions" (\ref{eq.125a}), (\ref{eq.125b}) satisfy the following systems of "eigenequations": 
\begin{eqnarray}
q^{i}\star\varrho_{\overline{q}_{1}\overline{q}_{2}}=q_{1}^{i}
\varrho_{\overline{q}_{1}
\overline{q}_{2}}, &\quad& \varrho_{\overline{q}_{1}\overline{q}_{2}}\star q^{i}=
q_{2}^{i}\varrho_{\overline{q}_{1}\overline{q}_{2}},\label{eq.142}\\
p_{i}\star\varrho_{\overline{p}_{1}\overline{p}_{2}}=p_{1i}\varrho_{\overline{p}_{1}
\overline{p}_{2}}, &\quad& \varrho_{\overline{p}_{1}\overline{p}_{2}}\star p_{i}=
p_{2i}\varrho_{\overline{p}_{1}\overline{p}_{2}}.\label{eq.143}
\end{eqnarray}
This is an obvious counterpart of the operator "eigenequations":
\begin{eqnarray}
\mathbf{q}^{i}|\overline{q}_{1}\rangle \langle \overline{q}_{2}|=q_{1}^{i}|
\overline{q}_{1}\rangle \langle \overline{q}_{2}|,&\quad& |\overline{q}_{1}\rangle 
\langle \overline{q}_{2}|\mathbf{q}^{i}=q_{2}^{i}|\overline{q}_{1}\rangle 
\langle \overline{q}_{2}|,\label{eq.144}\\
\mathbf{p}_{i}|\overline{p}_{1}\rangle \langle \overline{p}_{2}|=p_{1i}|
\overline{p}_{1}\rangle \langle \overline{p}_{2}|,&\quad& |\overline{p}_{1}\rangle 
\langle \overline{p}_{2}|\mathbf{p}_{i}=p_{2i}|\overline{p}_{1}\rangle 
\langle \overline{p}_{2}|.\qquad\label{eq.145}
\end{eqnarray}
In particular, we have the commutator "eigenequations": 
\begin{eqnarray}
\left[\mathbf{q}^{i},|\overline{q}_{1}\rangle \langle \overline{q}_{2}|\right]&=&
\left(q_{1}^{i}-q_{2}^{i}\right)|\overline{q}_{1}\rangle \langle 
\overline{q}_{2}|,
\label{eq.146}\\
\left[\mathbf{p}_{i},|\overline{p}_{1}\rangle \langle \overline{p}_{2}|\right]&=&
\left(p_{1i}-p_{2i}\right)|\overline{p}_{1}\rangle \langle \overline{p}_{2}|.
\label{eq.147}
\end{eqnarray}
The quotation marks at "eigenequations" are just due to the fact
that here we are outside the rigorous Hilbert space framework, but do not find it possible here to get into the complicated framework of rigged
Hilbert spaces. 

Similarly, if the basic states $\left|i\right\rangle$
in (\ref{eq.122}) are eigenstates of the Hermitian operator $\mathbf{A}$,
\begin{equation}
\mathbf{A}\left|i\right\rangle  =a_{i}\left|i\right\rangle,
\label{eq.148}
\end{equation}
where for simplicity the non-degenerate case is assumed, then, of course,
the operator eigenequations hold: 
\begin{eqnarray}
\mathbf{A}|i\rangle \langle j|&=&a_{i}|i\rangle \langle j|,\qquad |i\rangle \langle j|\mathbf{A}=a_{j}|i\rangle \langle j|,
\label{eq.149}\\
\left[\mathbf{A},|i\rangle \langle j|\right]&=&\left(a_{i}-a_{j}\right)|i\rangle \langle j|,\label{eq.150}
\end{eqnarray}
as well as their phase-space expressions in terms of the star-product.

The action of operators $\mathbf{A}$ on phase-space functions $A$
is represented by the following formula: 
\begin{equation}
\left(\mathbf{A}\psi\right)\left(\overline{q}\right)=\frac{1}
{\left(2\pi\hbar\right)^{n}}\int\exp\left(\frac{i}{\hbar}\overline{p}
\cdot\left(\overline{q}-\overline{q}'\right)\right)A\left(\frac{1}{2}
\left(\overline{q}+\overline{q}'\right),p\right)\psi\left(\overline{q}'
\right)d_{n}\overline{q}'d_{n}\overline{p}.\label{eq.151}
\end{equation}
Let us notice that this formula is correctly affine, i.e., translation-invariant,
because $\left(\overline{q}-\overline{q}'\right)$ represents the
translation vector from $\overline{q}$ to $\overline{q}'$, and $\left(\overline{q}+\overline{q}'\right)/2$
is the centre of the segment between $\overline{q}$ and $\overline{q}'$.
They are both coordinate-independent concepts.

Substituting the trigonometric representation:
\begin{equation}
\psi\left(q\right)=
f\left(q\right)\exp\left(\frac{i}{\hbar}S\left(q\right)\right)=
\sqrt{D\left(q\right)}\exp\left(\frac{i}{\hbar}S\left(q\right)\right),
\label{eq.152}
\end{equation}
we obtain in the limit $\hbar\rightarrow 0$ the following expansion
for (\ref{eq.151}):
\begin{equation}
\left(\mathbf{A}\psi\right)\left(q\right)\approx A\left(q^{i},\frac{\partial S}{\partial q^{i}}\right)\psi\left(q\right)+\frac{\hbar}{i}\left(\pounds_{v}f\right)
\exp\left(\frac{i}{\hbar}S\left(q\right)\right),\label{eq.153/153}
\end{equation}
and terms of the higher order in $\hbar$ are neglected. In this formula
$\pounds_{v}$ denotes the Lie derivative operator with respect to
the vector field $v\left[A,S\right]$ of velocities, 
\begin{equation}
v^{i}=\frac{\partial A}{\partial p_{i}}\left(q^{j},\frac{\partial S}
{\partial q^{j}}\right);\label{eq.154/154}
\end{equation}
obviously velocities in the sense of the Hamiltonian vector field
$X\left[A\right]$ generated by $A$, 
\begin{equation}
X\left[A\right]=\frac{\partial A}{\partial p_{i}}\frac{\partial}{\partial q^{i}}-\frac{\partial A}{\partial q^{i}}\frac{\partial}{\partial p_{i}}.\label{eq.155/155}
\end{equation}
More precisely, $X\left[A\right]$ is tangent to the Lagrangian manifold
$\mathfrak{M}_{S}\subset\mathbb{R}^{2n}$ given by equations:
\begin{equation}
p_{i}=\frac{\partial S}{\partial q^{i}},\qquad i=1,...,n.\label{eq.156/156}
\end{equation}
Then we take the restriction of $X_{A}$ to $\mathfrak{M}_{S}$,
\begin{equation}
X\left[A,S\right]:=\left.X\left[A\right]\right|_{\mathfrak{M}_{S}},
\label{eq.157/157}
\end{equation}
and project this vector field on $\mathfrak{M}_{S}$ to the configuration
space $\mathbb{R}^{n}$ (roughly speaking, the space of $q$-variables),
and just obtain the vector field $v\left[A,S\right]$ with components
$v^{i}$ given by (\ref{eq.154/154}). It must be stressed that geometrically
$f$ is not a scalar, but scalar density of weight $1/2$
\cite{Mack, JJS-Heis, JJS-phase}, therefore, its Lie derivative is given by 
\begin{equation}
\pounds_{v}f=v^{i}\frac{\partial f}{\partial q^{i}}+\frac{1}{2}\frac{\partial v^{i}}{\partial q^{i}}f.\label{eq.158/158}
\end{equation}

The Born probability density $D=f^{2}$ is a scalar density of weight
one, therefore,
\begin{equation}
\pounds_{v}D=v^{i}\frac{\partial D}{\partial q^{i}}+\frac{\partial v^{i}}{\partial q^{i}}D=\frac{\partial}{\partial q^{i}}\left(Dv^{i}\right).\label{eq.159/159}
\end{equation}
It follows in particular that in the classical limit Schr\"{o}dinger equation,
\begin{equation}
i\hbar\frac{\partial\psi}{\partial t}=\mathbf{H}\psi,\label{eq.160/160}
\end{equation}
results, as is well known from the standard treatment, in the system
consisting of the Hamilton-Jacobi equation for $S$ and continuity
equation:
\begin{eqnarray}
\frac{\partial S}{\partial t}+H\left(q^{i},\frac{\partial S}{\partial q^{i}},t\right) & = & 0,\label{eq.161/161}\\
\frac{\partial D}{\partial t}+\frac{\partial}{\partial q^{i}}j^{i} & = & 0,
\label{eq.162/162}
\end{eqnarray}
where $j^{i}$ is the contravariant vector density of weight one, physically
interpreted as the probability current, 
\begin{equation}
j^{i}=Dv\left[H,S\right]^{i}=D\frac{\partial H}{\partial p_{i}}\left(q,\frac{\partial S}{\partial q}\right).\label{eq.163/163}
\end{equation}
Geometrically, (\ref{eq.162/162}) may be interpreted as
\begin{equation}
\frac{\partial D}{\partial t}+\pounds_{v\left[H,S\right]}D=0.\label{eq.164/164}
\end{equation}

There are some discussions concerning the standard demonstration of
the classical limit, based on the $\hbar\rightarrow 0$ limit transition.
There are physicists (e.g., Frank Schroeck) who claim that the
procedure is incorrect or just wrong. Simply $\hbar$ is a constant,
e.g., one in the natural system of units. Of course, one must be careful
with such formal manipulations with constants, e.g., $\hbar\rightarrow 0$ and $c\rightarrow \infty$ (quasiclassical and non-relativistic limits). And
there are various misunderstandings when the transition procedure
is performed automatically. Some kind of "feeling" and intuition
is necessary. Physically, the procedure $\hbar\rightarrow 0$ is a
shorthand for dealing with large quantum numbers and quickly oscillating
wave functions, like in the method of stationary phase. There are
certain advantages of the formal manipulations with physical constants
as free parameters. It is not excluded that this procedure may be
somehow related to fundamental problems and interpretation questions
like the Anthropic Principle.

One point must be stressed here. For simplicity we worked in arithmetic
spaces $\mathbb{R}^{n}$, $\mathbb{R}^{2n}$ respectively as the configuration
and phase spaces. This was to avoid the crowd of symbols at the very
beginning and in some analytical formal transitions. But for the completeness
and "honesty" some comments are necessary. As configuration space
it is not $\mathbb{R}^{n}$ but rather some affine space $\left(M,V,\rightarrow\right)$
that should be used. Here $M$ is an underlying set and $V$ is the
$n$-dimensional real linear space if translations in $M$. The arrow
symbol denotes the operation which to any pair of points $x,y\in M$
assigns the vector $\overrightarrow{xy}\in V$ originating at $x\in M$
and terminating at $y\in M$, i.e., the vector of translation which
carries over $x$ into $y$. When the time unit is fixed and the
time axis is identified with $\mathbb{R}$, then $V$ may be also
identified with the space of virtual velocities. Its dual space $V^{\ast}$,
i.e., the space of $\mathbb{R}$-linear functions on $V$ is identified
with the space of canonical momenta. The Newton space of states is
$P_{N}=M\times V$, the space of positions and velocities. The phase
space, i.e., the space of Hamiltonian states is $P=M \times V^{*}$. Both $P_{N}$ and $P$ are affine spaces with translation spaces respectively $V \times V$ and $V \times V^{*}$. In realistic models some Euclidean structure in $M$ is also fixed, i.e., some symmetric and positively definite metric tensor $g \in V^{*} \otimes V^{*}$ is distinguished. What in $\mathbb{R}^{n}$ was denoted as $\overline{u} \cdot \overline{v}$, i.e., the standard scalar product, is now replaced by
\begin{equation}\label{eq.165/165}
\left(\overline{u} | \overline{v} \right):= g \left(\overline{u}, \overline{v} \right) = g_{ij}u^{i}v^{j}.
\end{equation}
Obviously, this reduces to the standard arithmetic expression $\overline{u} \cdot \overline{v}$ in $\mathbb{R}^{n}$ when orthonormal basis is used, i.e., such one that $g_{ij}=\delta_{ij}$. The symplectic form (\ref{eq.46}) on $\mathbb{R}^{2n} \simeq \mathbb{R}^{n} \times \mathbb{R}^{n}$ was expressed analytically in terms of the scalar product on $\mathbb{R}^{n}$. In that the formula (\ref{eq.46}) is an artificial and misleading artifact. As a matter of fact, the self-dual linear space $V \times V^{*}$ is endowed with the canonical symplectic two-form $\Gamma$ which has nothing to do with the metric tensors, i.e., scalar product on $V$, and is completely independent on any metrical concepts in $V$. Metric in $V$ is superfluous for the construction of $\Gamma$ as a skew-symmetric metric bilinear form on $V \times V^{*}$. Namely, the evaluation of $\Gamma$ on the pair of vectors $\overline{w}_{1} = \left(\overline{v}_{1}, \underline{p}_{1}\right) \in V \times V^{*}$ and $\overline{w}_{2} = \left(\overline{v}_{2}, \underline{p}_{2}\right) \in V \times V^{*}$ is given by
\begin{eqnarray}
\Gamma\left(\overline{w}_{1}, \overline{w}_{2}\right) &=& \Gamma\left(\left(\overline{v}_{1}, \underline{p}_{1}\right), \left(\overline{v}_{2}, \underline{p}_{2}\right)\right)=\label{eq.166/166} \\
&=& \underline{p}_{1}\left(\overline{v}_{2}\right) - \underline{p}_{2}\left(\overline{v}_{1}\right) = \left\langle \underline{p}_{1},\overline{v}_{2} \right\rangle - \left\langle \underline{p}_{2},\overline{v}_{1} \right\rangle, \nonumber
\end{eqnarray}
where in the last expression, $\left\langle \underline{p},\overline{v} \right\rangle$ is a commonly used evaluation of the linear function $\underline{p} \in V^{*}$ on the vector $\overline{v} \in V$. One must remember about all that when analytically the $\mathbb{R}^{n}$-formulas like (\ref{eq.46}) are used. The point is that in this analytical $\mathbb{R}^{n}$-language the distinction, or rather very deep logical difference, between linear spaces $V$ and $V^{*}$ is obscured.

Another our misleading silence committed for simplification and brevity is one concerning physical dimensions. Obviously, the absolute time is not the real axis, it is some one-dimensional affine space $T$ with the one-dimensional linear space of translations $\Lambda$. Then the space of virtual velocities is not just the space of translations $V$, but rather $\mathcal{V} = V \otimes \Lambda^{*}$. $\mathcal{V}$ becomes identified with $V$ when some basic element, i.e., the time unit is chosen in $\Lambda$ (and dually --- in $\Lambda^{*}$). Obviously $\mathcal{V}^{*}$, the dual space of $\mathcal{V}$ is isomorphic with $V^{*} \otimes \Lambda$. The next point is that strictly speaking, the metric tensor as a bilinear symmetric form on $V$ should take values not just in $\mathbb{R}$, but rather in some one-dimensional space $L \otimes L$, where $L$ is also one-dimensional space of lengths. Again, when some basis in $L$ is fixed ("metre" or "centimetre"), $g$ may be identified with an $\mathbb{R}$-valued symmetric bilinear form on $V$. The action function $S$ in the Hamilton-Jacobi equation is roughly considered as an $\mathbb{R}$-valued scalar. But it has just the physical dimension of action, thus, it rather takes its values in some one-dimensional space of action $A$. Similarly, the Cartan differential one-form $p_{i}dq^{i}$ should be rather $A$-valued than $\mathbb{R}$-valued. Then the quantity $p_{i}(dq^{i}/dt)$ should have the physical dimension of the action divided by time, i.e., energy. This would mean that $p$ should be an element of the one-dimensional space $A \otimes V^{*} \otimes \Lambda$. We shall not develop this motive of physical dimension. Unlike what it may seem, it is neither academic nor trivial. Obviously, the Planck constant $\hbar$ is the basic element of $A$ distinguished by physics and discovered in the analysis of quite concrete and important physical phenomena. Exponential function $\exp$ may be defined only for the purely numerical, dimension-less arguments. And it is just here where the Planck constant explicitly occurs in expressions like (\ref{eq.152}), e.g.,
\begin{equation}
\sqrt{D} \exp\left(\frac{i}{\hbar} \ S\right), \qquad 
\exp\left(\frac{i}{\hbar} \ p_{a}q^{a}\right).\label{eq.153/167}
\end{equation}

Some academic questions may appear like whether $\hbar$ is to appear at all explicitly in physical formulas. Why not to write simply
\begin{equation}
\sqrt{D} \exp\left(i\sigma \right), \qquad \exp\left(i k_{a}q^{a}\right), \label{eq.154/168}
\end{equation}
where $\overline{k}$ denotes the wave vector? The point is that it is $p_{a} = \hbar k_{a}$, the "eigenvalue" of $(\hbar/i)(\partial/\partial q^{a})$ (by abuse of language; one deals here with continuous spectrum and non-$L^{2}$ wave functions) that is directly interpretable in particle phenomena and experiments as the linear momentum. Situation is slightly different in the case of discrete spectrum, e.g., when one deals with wave functions of the planar rotator,
\begin{equation}
\psi_{n}(\varphi) \simeq \exp\left(i n \varphi \right), \label{eq.155/169}
\end{equation}
where $\varphi$ denotes the angular variable, $n$ is an integer, $n \in \mathbb{Z}$. Obviously, (\ref{eq.155/169}) is the eigenfunction of the rotational angular momentum. It corresponds to the eigenvalue $n \hbar$, because
\begin{equation}
\frac{\hbar}{i}\frac{\partial}{\partial \varphi} \ \psi_{n} = \hbar n \psi_{n}. \label{eq.156/170}
\end{equation}
Obviously, it would be strange and non-natural to introduce explicitly the Planck constant $\hbar$ to the exponent of (\ref{eq.155/169}),
\begin{equation}
\psi_{n} \simeq \exp \left(\frac{i}{\hbar}(n \hbar) \varphi\right). \label{eq.157/171}
\end{equation}

The same problem occurs in eigenfunctions of the orbital angular momentum in three dimensions. Nevertheless, it is just $j_{n}= n \hbar$ that is physically interpretable as angular momentum and in this meaning it does occur in all formulas where the classically interpretable angular momentum appears. Of course, there are some new problems in quasiclassical analysis, because the formal automatism of $\hbar \rightarrow 0$ does not work directly. What does work are large quantum numbers and quickly oscillating eigenfunctions. The first step of analysis consists in the trick of approximate replacing of discrete spectrum by continuous one. Namely, one takes superpositions of the form:
\begin{equation}
\psi(\varphi)= \sum_{n} c_{n} \exp \left(in \varphi \right), \label{eq.158/172}
\end{equation}
where the discrete function $\mathbb{Z} \ni n \mapsto c_{n} \in \mathbb{C}$ is essentially concentrated in some range 
\begin{equation}
n_{0} - \Delta n \ll n \ll n_{0} + \Delta n, \label{eq.159/173}
\end{equation}
where
\begin{equation}
n_{0} \gg \Delta n \gg 1, \label{eq.160/174}
\end{equation}
and is there slowly-varying, 
\begin{equation}
\frac{\left|c_{n+1} - c_{n}\right|}{\left|c_{n}\right|} \ll 1. \label{eq.161/175}
\end{equation}
Then, as it is well-known from the theory of Fourier analysis, the Fourier series (\ref{eq.158/172}) may be in a good approximation replaced by the integral expression:
\begin{equation}
\psi(\varphi) = \int c(k) \exp(ik \varphi)dk, \label{eq.162/176}
\end{equation}
where at the lattice points from the range (\ref{eq.159/173}) $c(k)$ equals $c_{n}$ and is regularly varying between the lattice points, e.g., linearly. Then the functions $\psi(\varphi)$ are essentially concentrated in a small domain of the variable $\varphi$ and "forget" the circular topology of the $\varphi$-variable. They may be interpreted as "well-behaving at infinity" functions on $\mathbb{R}$, and apparently the topology and geometry of $T^{1} \simeq {\rm SO}(2, \mathbb{R}) \simeq {\rm U}(1)$ is replaced by that of $\mathbb{R}$. And then, after the substitution $k=p / \hbar$, everything reduces to the problem on $\mathbb{R}$. And further on
the previous reasoning may be repeated. The same is true for $n$ circular degrees of freedom, i.e., for the toroidal configuration space
\begin{equation}\label{eq.163}
T^{n}\simeq {\rm SO}(2, \mathbb{R})^{n}\simeq {\rm U}(1)^{n}.
\end{equation}
And then, after reducing the problem to $\mathbb{R}^{n}$, one can repeat the above $\hbar \rightarrow 0$ quasiclassical asymptotic procedure.

To complete the above remarks about $\hbar \rightarrow 0$ asymptotics, let us remind the limit formulas for the Weyl-Wigner-Moyal-Ville star product and for the quantum Poisson bracket. One can show that 
\begin{equation}\label{eq.164}
\lim_{\hbar \rightarrow 0} A \star B=AB,
\end{equation}
i.e., in the limit one obtains the local pointwise product. It is assumed here that $A$, $B$ themselves do not depend on $\hbar$. More generally, one can construct the asymptotic series for $A \star B$ as a function of $\hbar$. The expansion up to first-order terms reads
\begin{equation}\label{eq.165}
A\star B=AB+\frac{i\hbar}{2}\{A,B\}+ \cdots,
\end{equation}
where, obviously, $\{A,B\}$ is the usual Poisson bracket of the phase-space functions $A$, $B$. Therefore, for the quantum Poisson bracket expressed in the phase-space terms,
\begin{equation}\label{eq.166}
\{A,B\}_{\rm quant}:=\frac{1}{i\hbar}(A\star B - B\star A),
\end{equation}
one obtains, as expected that
\begin{equation}\label{eq.167}
\lim_{\hbar \rightarrow 0}\{A,B\}_{\rm quant}=\{A,B\}.
\end{equation}
The formulas (\ref{eq.137}), (\ref{eq.138}), (\ref{eq.140}) seem to tell us that on the classical level the pure states are represented by singular probability distributions concentrated on $n$-dimensional submanifolds of the $2n$-dimensional classical phase space. These supports are very special $n$-dimensional submanifolds, namely, Lagrangian submanifolds\cite{AbrMar, Arnold, JJS-Heis, JJS-phase}. 
Evaluation of the symplectic two-form $\Gamma$ on any pair of their tangent vectors at any point does vanish. This is the classical expression of uncertainty relations. There is also a close relationship to the optical-mechanical analogy.

Let us notice however that for the coherent states given by (\ref{eq.113}) we obtain Dirac distributions concentrated on single points,
\begin{equation}\label{eq.168}
\lim_{\hbar \rightarrow 0}E_{\left(\overline{\xi}, \overline{\pi}\right)}(\overline{q}, \overline{p})=\delta\left(\overline{q}-\overline{\xi}\right)
\delta\left(\overline{p}-\overline{\pi}\right),
\end{equation}
therefore, something completely different. The point is however that, as mentioned, $\hbar \rightarrow 0$ is a shorthand for dealing with large quantum numbers and quickly oscillating wave functions. But (\ref{eq.113}) is a description of oscillatory ground states, therefore, something with extremely "small quantum numbers". Therefore, it seems that (\ref{eq.168}) is just an example of the situation where the $\hbar \rightarrow 0$ formal transition is not justified.

It is interesting and instructive to express the Weyl-Wigner-Moyal-Ville formulation and related problems in terms of the central extension \cite{Bargmann} of the classical phase space and of the Newton subgroup of the Galilei group.

We add an additional one dimension of the ${\rm U}(1)$-group, with the angular coordinate $\varphi$ of the dimension of action, or the dimension-less $\xi:=\varphi/\hbar$. Then we define
\begin{equation}\label{eq.169}
Z\left\{\xi; \overline{\alpha}, \overline{\nu}\right\}=
Z\left(\varphi;  \overline{\alpha}, \overline{\nu}\right):=
\exp \left(\frac{i}{\hbar}\varphi\right)\mathcal{W}\left( \overline{\alpha}, \overline{\nu}\right)=\exp \left(i\xi\right)
\mathcal{W}\left(\overline{\alpha}, \overline{\nu}\right)
\end{equation} 
with the composition rule
\begin{eqnarray}
Z\left\{\xi_{1}; \overline{\alpha}_{1}, \overline{\nu}_{1}\right\}Z\left\{\xi_{2}; \overline{\alpha}_{2}, \overline{\nu}_{2}\right\}&=&\exp \left(\frac{i}{\hbar}(\varphi_{1}+\varphi_{2}+
m\overline{\nu}_{1}\cdot\overline{\alpha}_{2})\right)\nonumber\\
&&\mathcal{W}\left(\overline{\alpha}_{1}+\overline{\alpha}_{2}, \overline{\nu}_{1}+\overline{\nu}_{2}\right).\label{eq.170}
\end{eqnarray} 
It is again convenient to take the "elegantly" ordered operations:
\begin{equation}\label{eq.171}
Z\left\{\xi; \overline{\alpha}, \overline{\nu}\right\}=
Z\left(\varphi; \overline{\alpha}, \overline{\nu}\right):=
\exp\left(\frac{i}{\hbar}\left(\varphi+\frac{m}{2}\overline{\nu}
\cdot\overline{\alpha}\right)\right)U(\overline{\alpha})V(\overline{\nu}).
\end{equation}
Then one obtains
\begin{eqnarray}
Z(\varphi_{1}; \overline{\alpha}_{1}, \overline{\nu}_{1})Z(\varphi_{2}; \overline{\alpha}_{2}, \overline{\nu}_{2})&=&Z\left( \varphi_{1}+\varphi_{2}\right.\label{eq.172}\\
&+&\left.\frac{m}{2}\Gamma \left((\overline{\alpha}_{1}, \overline{\nu}_{1}), (\overline{\alpha}_{2}, \overline{\nu}_{2}\right)); \overline{\alpha}_{1}+\overline{\alpha}_{2}, \overline{\nu}_{1}+\overline{\nu}_{2}\right),\nonumber
\end{eqnarray}
or, using the systematic $\mathbb{R}^{2n}$-language,
\begin{equation}\label{eq.173}
Z(\varphi_{1}; \overline{u}_{1})Z(\varphi_{2}; \overline{u}_{2})=
Z\left( \varphi_{1}+\varphi_{2}+\frac{m}{2}\Gamma (\overline{u}_{1}, \overline{u}_{2}); \overline{u}_{1}+\overline{u}_{2}\right).
\end{equation} 
Let us also quote the triple rule corresponding to the associativity,
\begin{eqnarray}
Z(\varphi_{1};u_{1})Z(\varphi_{2}; u_{2})Z(\varphi_{3};u_{3})&=&
Z\left( \varphi_{1}+\varphi_{2}+\varphi_{3}\right.\nonumber\\
&+&\left.\frac{m}{2}
\Gamma (\overline{u}_{1}, \overline{u}_{2}, \overline{u}_{3}); \overline{u}_{1}+\overline{u}_{2}+\overline{u}_{3}\right),\label{eq.174}
\end{eqnarray} 
where the following co-cycle notation is used:
\begin{equation}\label{eq.175}
\Gamma (\overline{u}_{1}, \overline{u}_{2}, \overline{u}_{3})=\Gamma (\overline{u}_{2}, \overline{u}_{3})+\Gamma (\overline{u}_{3}, \overline{u}_{1})+\Gamma (\overline{u}_{1}, \overline{u}_{2}).
\end{equation} 
Let us also notice that
\begin{equation}\label{eq.176}
Z\left(\varphi, \overline{u}\right)^{-1}=Z\left(-\varphi, -\overline{u}\right).
\end{equation} 

On the manifold of the centrally extended group we can introduce the convolution of functions in the sense of the above multiplication rule. Then,
\begin{equation}\label{eq.177}
(A \ast B)\left(\psi, \overline{w}\right)=\int A (\varphi, \overline{u})B \left( \psi-\left(\varphi+\frac{m}{2}\Gamma(\overline{u}, \overline{w}) \right); \overline{w}-\overline{u}\right)\frac{d\varphi}{2\pi}d_{2n}\overline{u}.
\end{equation} 
The $(2\pi)$-denominator is not essential of course; this is simply normalization of the "volume" of ${\rm U}(1)$ to unity.

Let us now restrict the rule to the subspace of eigenfunctions of $(1/i)\partial/ \partial \varphi$ with the eigenvalue $n$,
\begin{equation}\label{eq.178}
A(\varphi, \overline{u})=\exp (in \varphi)a(\overline{u}), \qquad B(\varphi, \overline{u})=\exp (in \psi)b(\overline{u}).
\end{equation} 
Their convolution belongs to the same subspace, because for two such $n$-func\-tions,
\begin{equation}\label{eq.179}
A(\varphi, \overline{u})=\exp (in \varphi)a(\overline{u}), \quad B(\varphi, \overline{u})=\exp (in \varphi)b(\overline{u}),
\end{equation} 
we have
\begin{equation}\label{eq.180}
(A\ast B)(\psi, \overline{w})= \exp \left(in \psi\right)\int 
\exp \left( -in \frac{M}{2}\Gamma(\overline{u}, \overline{w})\right) a(\overline{u})b(\overline{w}-\overline{u})d\overline{u},
\end{equation} 
where $M$ denotes the particle mass. 

If instead of a fixed value of $n$ we take the whole series,
\begin{equation}\label{eq.181}
A(\varphi, \overline{u})=\sum_{n\in \mathbb{Z}} \exp (in\varphi)a_{n}(\overline{u}), \qquad B(\varphi, \overline{u})=\sum_{k} \exp (ik\psi)b_{k}(\overline{u}),
\end{equation}
we obtain
\begin{equation}\label{eq.182}
(A\ast B)(\psi, \overline{w})=\sum_{n\in \mathbb{Z}}\exp (in \psi)\left( a_{n}\underset{n,\Gamma}{\ast}b_{n}\right)(\overline{w}),
\end{equation}
where
\begin{equation}\label{eq.183}
\left( a\underset{n,\Gamma}{\ast}b\right)(\overline{w})=\int \exp \left( -in \frac{M}{2}\Gamma(\overline{u}, \overline{w})\right)a(\overline{u})b(w-\overline{u})d_{2n}\overline{u}.
\end{equation}

Let us stress again that in the formulas from (\ref{eq.177}) to (\ref{eq.183}) 
one is dealing with the usual group convolution, not with the Weyl-Wigner-Moyal-Ville star-product. 

The idea here is the Fourier analysis for functions on $\mathbb{R}^{2n}$ and the corresponding splitting into invariant subspaces. Nevertheless, the parallel reasoning may be performed for the star-product. Having other problems in mind, we shall not discuss this one here; it is postponed to subsequent papers.

Let us consider some generalization. In the above presentation of the Moyal treatment it was essential that $\mathbb{R}^{2n}$, or, more correctly, $P=V \times V^{\ast}$, being a linear space, was an Abelian group under the usual addition of vectors. So, let us take instead of $\mathbb{R}^{2n}$ some locally compact Abelian group $G$. Then one can use generalized Fourier analysis based on Pontryagin duality. In this case all irreducible unitary representations are one-dimensional. Being ${\rm U}(1)$-valued functions on $G$, they form the set closed under the pointwise multiplication. They form the dual group of $G$, denoted by $\widehat{G}$ and referred to as the character group \cite{Rudin, Loomis}. The term is correct, because irreducible unitary representations are now one-dimensional and just coincide in a trivial way with their characters, i.e., traces of the representation matrices.

The group operation in $\widehat{G}$ is meant
as the pointwise multiplication of functions on $G$. In the other
words, the elements of $\widehat{G}$ are continuous homomorphisms
of $G$ into the group ${\rm U}(1)=\left\{z\in\mathbb{C}:\left|z\right|=1\right\}$,
the multiplicative group of complex numbers of modulus one. If $G$ is
compact and Abelian, then $\widehat{G}$ is discrete, and the Peter-Weyl
series expansion becomes a generalized Fourier
series. If $G$ is non-compact, one obtains generalized Fourier transforms
and direct integrals of family of one-dimensional spaces.

According to the well-known Pontryagin theorem, the dual of $\widehat{G}$,
e.g., the second dual $\widehat{\widehat{G}}$ of $G$, is canonically
isomorphic with $G$ itself. This resembles the relationship between
duals of finite-dimensional linear spaces, $\left(V^{*}\right)^{*}\simeq V$.

The Fourier transform $\widehat{\psi}:\widehat{G}\rightarrow\mathbb{C}$ of $\psi:G\rightarrow\mathbb{C}$ is defined as follows:
\begin{equation}
\widehat{\psi}(\chi)=\int\overline{\left\langle \chi|g\right\rangle }\psi(g)dg=\int\left\langle \chi|g\right\rangle ^{-1}\psi(g)dg,\label{eq_1.137}
\end{equation}
where $dg$ denotes the integration element of the Haar measure
on $G$ and $\left\langle\chi|g\right\rangle$ is the evaluation
of $\chi\in\widehat{G}$ on $g\in G$; equivalently, in virtue of
Poincare duality, this is the evaluation of $g\in G\simeq\widehat{\widehat{G}}$
on $\chi\in\widehat{G}$. The inverse formula reads
\begin{equation}
\psi(g)=\int\left\langle \chi|g\right\rangle \widehat{\psi}(\chi)d\chi,\label{eq_1.138}
\end{equation}
where $d\chi$ denotes the element of Haar integration on $\widehat{G}$.
The formulas (\ref{eq_1.137}), (\ref{eq_1.138}) fix the synchronization
between normalizations of measures $dg$, $d\chi$. In principle,
these formulas are meant in the sense of $L^{1}$-spaces over $G$,
$\widehat{G}$, nevertheless, some more or less symbolic expressions
are also admitted for other functions, as shorthands for longer systems
of formulas. First of all, this concerns $\delta$-distributions,
just like in general situation of locally compact $G$.
Of course, the correct definition of distributions and operations on them
must be based on differential concepts, nevertheless, the Dirac distribution
itself (but not its derivatives) may be introduced in principle on
the basis of purely topological concepts, just like in the general
case. Let us notice that 
\begin{equation}
\psi(g)=\int d\chi\int dh\psi(h)\left\langle \chi|hg^{-1}\right\rangle.\label{eq_1.139}
\end{equation}
The order of integration here is essential! But, of course, one cannot
resist the temptation to change "illegally" this order and write
symbolically:
\begin{equation}
\psi(g)=\int dh\,\delta\left(hg^{-1}\right)\psi(h),\qquad\delta(x)=\int d\chi\left\langle \chi|x\right\rangle.\label{eq_1.140}
\end{equation}
If $G$ is discrete, then $\widehat{G}$ is compact (and conversely)
and the second integral is well defined, namely 
\begin{equation}
\delta(x)=\delta_{xe}
\begin{cases}
1, & \textrm{if }\ x=e,\\
0, & \textrm{if }\ x\neq e,
\end{cases}\label{eq_1.141}
\end{equation}
where, obviously, $e$ is the neutral element (identity) of $G$.
Then the first integral is literally true as a summation with the
use of Kronecker delta. But when obeying some rules, we may safely
use the formulas (\ref{eq_1.140}) also in the general case, when they
are formally meaningless. So, we shall always write
\begin{eqnarray}
\delta(g)&=&\int\left\langle \chi|g\right\rangle d\chi=\delta\left(g^{-1}\right), \label{eq_1.142a}\\
\delta(\chi)&=&\int\overline{\left\langle \chi|g\right\rangle}dg=
\int\left\langle \chi|g\right\rangle dg=\delta\left(\chi^{-1}\right),\label{eq_1.142b}\\
\int\delta(g)f(g)\, dg&=&f\left(e(G)\right),\label{eq_1.142c}\\
\int\delta(\chi)k(\chi)\, d\chi&=&
k\left(e(\widehat{G})\right),\label{eq_1.142d}
\end{eqnarray}
and $e(G)$, $e(\widehat{G})$ denote the units in $G$, $\widehat{G}$,
respectively. 

The convolution is defined by the usual formula:
\begin{equation}
\left(F\ast G\right)(g)=\int F(h)G\left(h^{-1}g\right)dh,
\end{equation}
and in the vector representation on $\mathbb{R}^{n}$ it simply becomes
\begin{equation}
\left(F\ast G\right)(u)=\int F(v)G(u-v)d_{n}v.
\end{equation}
But, obviously, the peculiarity of Abelian groups $G$ is that convolution
is a commutative operation, 
\begin{equation}
F*G=G*F.\label{eq_1.143}
\end{equation}
Obviously, Fourier transforms of convolution are pointwise products
of Fourier transforms, and conversely,
\begin{equation}
\left(F*G\right)^{\wedge}=\widehat{F}\widehat{G}.\label{eq_1.144}
\end{equation}
It is clear that just like in the general case, $\delta$-distribution
is the convolution identity, 
\begin{equation}
F*\delta=\delta*F=F.\label{eq_1.145}
\end{equation}

And now, we may be a bit more precise. Namely, let $U\subset\widehat{G}$
be some compact measurable subset of $\widehat{G}$, 
and let $L\left\{U\right\}$ denote
the linear subspace of functions (\ref{eq_1.138}) such that the Fourier
transform $\widehat{\psi}$ vanishes outside $U$ and is $L^{1}$-class.
Take the function $\delta\{U\}$ given by
\begin{equation}
\delta\{U\}(g):=\underset{U}{\int}\left\langle \chi|g\right\rangle d\chi.\label{eq_1.146}
\end{equation}
It is clear that $\delta\{U\}$ is a convolution identity of the subspace
$L\{U\}$. And now take an increasing sequence of subsets $V_{i}\subset\widehat{G}$ such that:
\begin{equation}\label{eq_1.146a}
V_{i}\supset V_{j}\quad\textrm{for}\quad i>j\,\qquad \underset{i}{\bigcup}\ V_{i}=\widehat{G}.
\end{equation}
It is clear that for any function $F\subset L^{1}(G)$ we have 
\begin{equation}
\underset{i\rightarrow\infty}{\lim}\delta\left\{ V_{i}\right\} *F=F,\label{eq_1.147}
\end{equation}
although the limit of the sequence $\delta\{V_{i}\}$ does not exist
in the usual sense of function sequences. However, it does exist in
an appropriately defined functional sense. So, by abuse of language,
we simply write:
\begin{equation}
\delta=\underset{i\rightarrow\infty}{\lim}\delta\{V_{i}\},\qquad\delta*F=F,
\label{eq_1.148}
\end{equation}
as a shorthand for the rigorous (\ref{eq_1.147}). 

Calculating formally the convolution of $\chi_{1},\chi_{2}\in\widehat{G}$,
we obtain 
\begin{equation}
\left\langle \chi_{1}*\chi_{2}|g\right\rangle =\delta\left(\chi_{1}\chi_{2}\!^{-1}\right)\left\langle \chi_{2}|g\right\rangle =\delta\left(\chi_{1}\chi_{2}\!^{-1}\right)\left\langle \chi_{1}|g\right\rangle,\label{eq_1.149}
\end{equation}
i.e., briefly 
\begin{equation}
\chi_{1}*\chi_{2}=\delta\left(\chi_{1}\chi_{2}\!^{-1}\right)\chi_{2}=
\delta\left(\chi_{1}\chi_{2}\!^{-1}\right)\chi_{1}.\label{eq_1.150}
\end{equation}

If $G$ is compact, i.e., $\widehat{G}$ is discrete, this is the
usual condition for irreducible idempotents. Similarity, we have the orthogonality/normalization condition 
\begin{equation}
\left(\chi_{1},\chi_{2}\right)=\delta\left(\chi_{1}\chi_{2}\!^{-1}\right)=
\begin{cases}
1, & \textrm{if }\ \chi_{1}=\chi_{2},\\
0, & \textrm{if }\ \chi_{1}\neq\chi_{2}.
\end{cases}\label{eq_1.151}
\end{equation}

If $G$ is not compact, i.e., $\widehat{G}$ is not discrete, then
both normalization and idempotence rules (\ref{eq_1.150}), (\ref{eq_1.151})
are meant symbolically, just like the corresponding rules for Dirac
distributions in $\mathbb{R}^{n}$: 
\begin{eqnarray}
\delta_{a}*\delta_{b} & = & \delta(a-b)\delta_{a}=\delta(a-b)\delta_{b},\label{eq_1.152a}\\
\left(\delta_{a},\delta_{b}\right) & = & \delta(a-b).\label{eq_1.152b} \end{eqnarray}

Obviously, $\delta_{a}(x):=\delta\left(x-a\right)$. Incidentally,
(\ref{eq_1.152a}), (\ref{eq_1.152b}) is just the special case of (\ref{eq_1.150}), (\ref{eq_1.151})
when $G=\mathbb{R}^{n}$ and the addition of vectors is meant as a
group operation.

The peculiarity of locally compact Abelian groups is that they offer
some analogies to geometry of the classical phase spaces and some
natural generalization of the Weyl-Wigner-Moyal formalism. Certain
counterparts do exist also in non-Abelian groups, especially compact
ones. However, they are radically different from the structures based
on Abelian groups. And in the non-compact case the analogy rather diffuses.

Finally, let us remind that just like in the classical Fourier analysis,
the Pontryagin Fourier transform is an isometry of $L^{2}(G)$ onto
$L^{2}\left(\widehat{G}\right)$,
\begin{equation}
\int\overline{A(g)}B(g)dg=\int\overline{\widehat{A}}(\chi)B(\chi)d\chi,
\label{eq_1.153}
\end{equation}
in particular, the Plancherel theorem holds 
\begin{equation}
\int\left|A(g)\right|^{2}dg=\int\left|\widehat{A}(\chi)\right|^{2}d\chi.
\label{eq_1.154}
\end{equation}

In physical applications, then, of course, the standard of Lebesgue measure is fixed,
\begin{equation}
\int f(x) d\mu (x) = \int f e^{1}\wedge \ldots \wedge e^{n}, \label{eq_2.101}
\end{equation}
where $\left(\ldots, e^{a}, \ldots\right)$ is an arbitrary orthonormal co-basis in $V^{\ast}$:
\begin{equation}
g = \delta_{ij} e^{i} \otimes e^{j}. \label{eq_2.102}
\end{equation}
In arbitrary coordinates, including  curvilinear ones, we have
\begin{equation}
\int f(x) d\mu (x) = \int f(x)\sqrt{\det \left[g_{ij}\right]} \ dx^{1} \ldots dx^{n}. \label{eq_2.103}
\end{equation}
The dual linear space $V^{\ast}$ parametrizes the dual group $\widehat{V}$ with the help of the standard covering homomorphism of the (additive) $\mathbb{R}$ onto (multiplicative) $U(1)$,
\begin{equation}
\mathbb{R} \ni \varphi \mapsto \exp (i \varphi) \in U(1), \label{eq_2.104}
\end{equation}
so, $\chi\left(\underline{k}\right) \in \widehat{V}$ is given by
\begin{equation}
\left\langle \chi\left(\underline{k}\right), \overline{x} \right\rangle= \exp \left(i \left\langle \underline{k} , \overline{x} \right\rangle \right), \label{eq_2.105}
\end{equation}
where, obviously, $\left\langle \underline{k} , \overline{x} \right\rangle$ is the evaluation of $\underline{k} \in V^{\ast}$ on $\overline{x} \in V$; analytically
\begin{equation}
\left\langle \underline{k} , \overline{x} \right\rangle = k_{a}x^{a}. \label{eq_2.106}
\end{equation}
Using the language of quantum momentum $\underline{p}= \hbar \underline{k}$, one writes also
\begin{equation}
\left\langle \chi\left[\underline{p}\right], \overline{x} \right\rangle= \exp \left(\frac{i}{\hbar} \left\langle \underline{p} , \overline{x} \right\rangle \right) = \exp \left(\frac{i}{\hbar} \ p_{a}x^{a}\right). \label{eq_2.107}
\end{equation}
The corresponding conventions of Fourier analysis, particularly popular in quantum mechanics, are as follows:
\begin{eqnarray}
f\left(\overline{x}\right) &=& \frac{1}{(2\pi)^{n}} \int \widehat{f}\left(\underline{k}\right) \exp \left(i \left\langle \underline{k}, \overline{x} \right\rangle \right)d_{n}k \nonumber\\
& =&\frac{1}{(2\pi \hbar)^{n}} \int \widehat{f}\left(\underline{p}\right) \exp \left(\frac{i}{\hbar} \left\langle \underline{p}, \overline{x} \right\rangle \right)d_{n}p,\label{eq_2.108a} \\
\widehat{f}\left(\underline{k}\right) = \widehat{f}\left[\underline{p}\right] &=& \int f\left(\overline{x}\right) \exp \left(-\frac{i}{\hbar} \left\langle \underline{p}, \overline{x} \right\rangle \right)d_{n}x.\label{eq_2.108b}
\end{eqnarray}
The convolution on $V$ is meant in the usual convention,
\begin{equation}
\left(A \ast B\right)\left(\overline{x}\right)= \int A(\overline{y}) B(\overline{x}-\overline{y})d\overline{y}. \label{eq_2.109}
\end{equation}
We have then the following rules,
\begin{eqnarray}
\chi\left(\underline{k}\right) \ast \chi\left(\underline{l}\right)&=& (2\pi)^{n}\delta\left(\underline{k}-
\underline{l}\right)\chi\left(\underline{k}\right)= (2\pi)^{n}\delta\left(\underline{k}-
\underline{l}\right)\chi\left(\underline{l}\right),\label{eq_2.110a}\\
\left(\chi\left(\underline{k}\right), \chi\left(\underline{l}\right)\right)&=& (2\pi)^{n}\delta\left(\underline{k}-\underline{l}\right),\label{eq_2.110b}\\
\chi\left[\underline{p}\right] \ast \chi\left[\underline{p'}\right]&=& (2\pi \hbar)^{n}\delta\left(\underline{p}-
\underline{p'}\right)\chi\left[\underline{p}\right]= (2\pi \hbar)^{n}\delta\left(\underline{p}-
\underline{p'}\right)\chi\left[\underline{p'}\right],\qquad \label{eq_2.111a}\\
\left(\chi\left[\underline{p}\right], \chi\left[\underline{p'}\right]\right)&=& (2\pi \hbar)^{n}\delta\left(\underline{p}-\underline{p'}\right),
\label{eq_2.111b}
\end{eqnarray}
rather unpleasant ones, because of the $(2\pi)^{n}$, $(2\pi \hbar)^{n}$-factors. But this has to do with the use of traditional symbols of analysis. If we remember that it is not $d_{n}k$, or $d_{n}p$, but rather $d_{n}k/(2\pi)^{n}$, $d_{n}p/(2\pi \hbar)^{n}$ that is a measure Fourier-synchronized with $d_{n}x$, that it is just $(2\pi )^{n}\delta\left(\underline{k}-\underline{l}\right)$ or $(2\pi \hbar)^{n}\delta\left(\underline{p}-\underline{p'}\right)$ that is to be interpreted as a "true Dirac delta", let us say $\Delta\left(\underline{k}-\underline{k'}\right)$, $\Delta\left(\underline{p}-\underline{p'}\right)$, respectively in the spaces of wave co-vectors and linear momenta.

There are various conventions concerning Fourier transforms and synchronization of measures on $G$, $\widehat{G}$, it is even stated in the book by Loomis \cite{Loomis}, that it is "an interesting and non-trivial problem".

In classical analysis one often prefers the "symmetric" convention:
\begin{eqnarray}
A\left(\overline{x}\right) &=&  \frac{1}{(2\pi)^{n/2}} \int \widehat{A}\left(\underline{k}\right) \exp \left(i \left\langle \underline{k}, \overline{x} \right\rangle \right)d_{n}\underline{k},\label{eq_2.112a} \\
\widehat{A}\left(\underline{k}\right) &=& \frac{1}{(2\pi)^{n/2}} \int A\left(\overline{x}\right) \exp \left(-i \left\langle \underline{k}, \overline{x} \right\rangle \right)d_{n}\overline{x}.\label{eq_2.112b}
\end{eqnarray}
An additional advantage of this convention is that the iteration of Fourier transformation results in the inversion (total reflection) of the original function, with respect to the origin:
\begin{equation}
\widehat{\widehat{A}{\,}}(x)=A(-x). \label{eq_2.113}
\end{equation}
And, roughly speaking, Gauss function is invariant under Fourier transformation. More precisely, we have
\begin{equation}
\mathcal{G}(\overline{x}) = \exp \left(-\frac{1}{2}\ \overline{x} \cdot \overline{x} \right), \qquad \mathcal{\widehat{G}}(\underline{k}) = \exp\left(-\frac{1}{2} \ \underline{k} \cdot \underline{k} \right), \label{eq_2.114}
\end{equation}
where the scalar product in $V$ is meant in the sense of metric $g \in V^{\ast} \otimes V^{\ast}$, and in $V^{\ast}$ --- under its contravariant inverse $g^{-1} \in V \otimes V$,
\begin{equation}\label{eq_2.114a}
\overline{x} \cdot \overline{x} = g(x,x)=g_{ij}x^{i}x^{j}, \qquad \underline{k} \cdot \underline{k} = g^{-1}(\underline{k},\underline{k})=g^{ij}x_{i}x_{j}. \end{equation}
If we identify $V=\mathbb{R}^{n}=V^{\ast}$, then the Gauss function is literally invariant under the Fourier transformation.

The Clebsch-Gordon formulas on $G$ have the following form:
\begin{eqnarray}
\chi(\underline{k})\chi(\underline{l})&=&\chi(\underline{k}+\underline{l}),
\label{eq_2.115a}\\ \chi\left[\underline{p}\right]\chi\left[\underline{p'}\right]&=&
\chi\left[\underline{p}+\underline{p'}\right],\label{eq_2.115b} \\
\chi(\underline{k})\chi(\underline{l})&=& \int \delta(\underline{k}+\underline{l}-
\underline{m})\chi(\underline{m})d_{n}\underline{m},\label{eq_2.115c} \\
\chi\left[\underline{p}\right]\chi\left[\underline{p'}\right]&=& \int \delta(\underline{p}+\underline{p'}-
\underline{\pi})\chi\left[\underline{\pi}\right]d_{n}\underline{\pi}.
\label{eq_2.115d}
\end{eqnarray}

Let us now fix some symbols concerning the compact case $T^{n}=U(1)^{n}$. Just like $\mathbb{R}^{n}$ is an analytical model of any $n$-dimensional linear space over reals, $T^{n}$ is parametrized by the system of angles $\left(\varphi^{1}, \ldots, \varphi^{n} \right)$ taken modulo $2\pi$, or uniquely, by the system of unimodular complex numbers $\left(\zeta^{1}, \ldots, \zeta^{n} \right)$, $\zeta^{a}= \exp \left(i \varphi^{a}\right)$. Sometimes the convention "modulo $1$" is accepted instead "modulo $2\pi$", i.e., one puts $\zeta^{a}= \exp \left(i2\pi \xi^{a}\right)$. This is often used when $T^{n}$ is realized as a quotient of $V$ modulo the "crystallographic lattice" generated freely by some fixed basis $\left(\ldots, e_{a}, \ldots\right)$ in $V$. Obviously, that discrete translation group is isomorphic with $\mathbb{Z}^{n}$. The parametrization modulo $2\pi$ is more popular in theory of Fourier series. Torus is compact and it is natural to take the Haar measure normalized to unity, as usual. If the multiple Fourier series on $T^{n}$ are meant in the convention
\begin{equation}
f(\overline{\varphi})= \sum_{\underline{m}\in\mathbb{Z}^{n}} \widehat{f}(\underline{m}) \exp \left(i \underline{m} \cdot \overline{\varphi} \right), \label{eq_2.116}
\end{equation}
then the inverse formula for coefficients $\widehat{f}$ reads
\begin{equation}
\widehat{f}(\underline{m})= \frac{1}{(2\pi)^{n}} \int f(\overline{\varphi}) 
\exp \left(-i \underline{m} \cdot \overline{\varphi} \right)d_{n}\overline{\varphi}. \label{eq_2.117}
\end{equation}
Concerning notation, analytical meaning of the expressions above is as follows:
\begin{equation}
\underline{m}=\left(m_{1}, \ldots, m_{n}\right) \in \mathbb{Z}^{n}, \qquad \overline{\varphi}= \left(\varphi^{1}, \ldots, \varphi^{n}\right)^{T}, \label{eq_2.118}
\end{equation}
contractions in exponents are given by
\begin{equation}
\underline{m}\cdot \overline{\varphi}=m_{a}\varphi^{a}= m_{1}\varphi^{1} + \cdots + m_{n}\varphi^{n}, \label{eq_2.119}
\end{equation}
and the range of variables $\varphi^{a}$ in the integration element 
\begin{equation}\label{eq_2.119a}
d_{n}\overline{\varphi}= d\varphi^{1} \ldots d\varphi^{n}
\end{equation}
is given by $[0,2\pi]$.

It is seen that the occurrence of factors $(2\pi)^{-n}$ is reciprocal to that in Fourier analysis on $\mathbb{R}^{n}$. This spoils the formal analogy, but suits the convention that the Haar volume of compact groups equals the unity. To save the analogy, we would have to replace (\ref{eq_2.108a})--(\ref{eq_2.108b}) by 
\begin{eqnarray}
f(\overline{x}) &=& \int \widehat{f}(\underline{k}) \exp \left(i\left\langle  \underline{k}, \overline{x}\right\rangle \right) d_{n}\underline{k},\label{eq_2.120a} \\
\widehat{f}(\underline{k}) &=& \frac{1}{(2\pi)^{n}}\int f(\overline{x}) \exp \left(-i\left\langle  \underline{k}, \overline{x}\right\rangle \right) d_{n}\overline{x},\label{eq_2.120b}
\end{eqnarray}
which, by the way, is sometimes used indeed, however, it is incompatible with some other customs of physicists and their taste.

Characters on $T^{n}$ are labelled by multi-indices $\underline{m}\in\mathbb{Z}^{n}$,
\begin{equation}
\left\langle \chi(\underline{m}), \zeta(\overline{\varphi})\right\rangle =\left(\zeta^{1}\right)^{m_{1}} \ldots \left(\zeta^{n}\right)^{m_{n}}= \exp \left(i \underline{m} \cdot \overline{\varphi}\right). \label{eq_2.121}
\end{equation}
The idempotence and independence property is literally satisfied, because $T^{n}$ is compact and $\mathbb{Z}^{n}$ is discrete:
\begin{eqnarray}
\chi(\underline{m}) \ast \chi(\underline{l})&=& \delta_{\underline{m}\underline{l}}\chi(\underline{m}) = \delta_{\underline{m}\underline{l}}\chi(\underline{l})\label{eq_2.122}\\
\chi(\underline{m}) \chi(\underline{l})&=& \chi(\underline{m}+ \underline{l}) \label{eq_2.123}\\
\left(\chi(\underline{m}), \chi(\underline{l})\right)&=& \delta_{\underline{m}\underline{l}}, \label{eq_2.124}
\end{eqnarray}
where, obviously, the multi-index Kronecker symbol $\delta_{\underline{m}\underline{l}}$ vanishes if $\underline{m} \neq \underline{l}$ (i.e., at least one component of $\underline{m}$ differs from the corresponding component of $\underline{l}$), and $\delta_{\underline{m}\underline{l}}=1$ when $\underline{m} = \underline{l}$. In other words
\begin{equation}
\delta_{\underline{m}\underline{l}}= \delta_{m_{1}l_{1}} \ldots \delta_{m_{n}l_{n}}. \label{eq_2.125}
\end{equation}
Concerning the "Clebsch-Gordon" rule (\ref{eq_2.123}), its representation reads
\begin{equation}
\chi(\underline{m}) \chi(\underline{l})= \sum_{\underline{\pi} \in \mathbb{Z}^{n}} \left(\underline{m} \ \underline{l} | \underline{\pi} \right) \left(\underline{m} \ \underline{l} | \underline{\pi} \right) \chi(\underline{\pi}), \label{eq_2.126}
\end{equation}
where
\begin{equation}
\left(\underline{m} \ \underline{l} | \underline{\pi} \right)= \delta_{\underline{m} + \underline{l} , \underline{\pi}}= \left(\underline{m} \ \underline{l} | \underline{\pi} \right)^{2}. \label{eq_2.127}
\end{equation}

Obviously, if we take as an arena of our physics the discrete group $\mathbb{Z}^{n}$, then its dual group $T^{n}$ is compact and continuous. Again the mentioned problems with squared delta-distribution appear.

Let us try to repeat the reasoning quoted above for $\mathbb{R}^{2n}$ (more precisely, as we saw, it was for $M\times V^{*}$). We assume the non-local star-product of complex functions on $G$:
\begin{equation}\label{eq.+1}
(A \star B)(g)=\int \mathcal{K}(g; g_{1}, g_{2})A(g_{1})B(g_{2})dg_{1}dg_{2}.
\end{equation}
It is assumed to be translationally invariant:
\begin{equation}\label{eq.+2}
U(g)(A \star B)=(U(g)A) \star (U(g)B),
\end{equation}
where $U(g)$ is the translation operator,
\begin{equation}\label{eq.+3}
(U(g)f)(h)=f\left(g^{-1}h\right).
\end{equation}
This implies that
\begin{equation}\label{eq.+4}
\mathcal{K}\left(g; g_{1}, g_{2}\right)=
K\left(g_{1}g^{-1}, g_{2}g^{-1}\right)=
K\left(g^{-1}g_{1}, g^{-1}g_{2}\right).
\end{equation}
Let us stress once more that $G$ is Abelian, otherwise there would be two kinds of translational invariance. Besides, we assume again that the star-product is to be associative:
\begin{equation}\label{eq.+5}
(A \star B)\star C=A\star (B\star C).
\end{equation}
This results in the following property of $K$:
\begin{equation}\label{eq.+6}
\int K(g_{1}, g)K\left(g_{2}g^{-1}, g_{3}g^{-1}\right)dg=\int 
K\left(g_{1}g^{-1}, g_{2}g^{-1}\right)K(g, g_{3})dg.
\end{equation}
The dual group of $G \times G$ is identical with $\widehat{G}\times\widehat{G}$, and the Fourier transform of $K$, denoted as usual by $\widehat{K}$, satisfies 
\begin{equation}\label{eq.+7}
\widehat{K}(\chi_{1}, \chi_{2}\chi_{3})\widehat{K}(\chi_{2}, \chi_{3})=
\widehat{K}(\chi_{1}, \chi_{2})\widehat{K}(\chi_{1}\chi_{2}, \chi_{3}).
\end{equation}
This holds, in particular, when $K$ is a two-character, i.e., if functions
\begin{equation}\label{eq.+8}
K(g, \cdot), \qquad K(\cdot, g)
\end{equation}
are characters, i.e., elements of $\widehat{G}$ for any $g\in G$.

If the mappings $g\mapsto K(g,\cdot)$,  $g\mapsto K(\cdot,g)$ are monomorphisms of $G$ into $\widehat{G}$, we say that the two-character $K$ is non-singular. Obviously, in general isomorphisms of $G$ onto $\widehat{G}$ do not exist, because $G$ and $\widehat{G}$ are, in general, non-isomorphic. Linear spaces as additive groups are important exceptions.

Now, let us consider the very special case which is analogous to the cotangent bundle, or more specifically, to the self-dual linear space $V\times V^{*}$, which carries the natural symplectic structure, as described above.

Namely, let us assume that the primary concept is the configuration space $G$ which is a locally compact Abelian group. Its dual group $\widehat{G}$ is assumed to be something like the set of canonical conjugate momenta. If $G=\mathbb{Z}^{n}$, physically $\mathbb{Z}^{3}$, the crystal lattice, then the canonical momenta are elements of the torus space $T={\rm U}(1)^{n}$, i.e., the direct product of $n$ circles. If we represent the elements of ${\rm U}(1)$ as
\begin{equation}\label{eq.+9}
\zeta=\exp (i \kappa)=\exp \left(\frac{i}{\hbar}\pi\right),
\end{equation}
then the conservation of momenta $\pi$ ("wave vectors" $\kappa$) will be damaged or rather modified in such a way that it is $\zeta$ that is conserved (e.g., in collision phenomena), but not necessarily $\pi$ or $\kappa$ (the $2\pi$-non-uniqueness of $\kappa$). These are so-called "Umklapp-processes". And conversely, if ${\rm U}(1)^{n}$ is the configuration space-torus, then the conjugate momenta are discrete, just "quantized", and are labelled by elements of $\mathbb{Z}^{n}$. Then the study of classical limit consists, as said above, of two steps: $(i)$ passing over from Fourier series to Fourier transforms (in some approximation, of course), and then $(ii)$ one considers the usual $\mathbb{R}^{2n}$-classical limit, formally as an asymptotics $\hbar \rightarrow 0$. The "Umklapp-processes" are physically observable.

It is interesting to quote some natural non-local products, i.e., natural kernels $K$ on the Abelian group $\mathcal{G}=G\times \widehat{G}$. Attention: now $\mathcal{G}$ is that was formerly denoted by $G$. Now $G$ is the "configuration space", and $\mathcal{G}=G\times \widehat{G}$ is the "phase space", do not confuse the symbols. The Kernels-two-characters $K$ are complex functions on $\mathcal{G}=G\times \widehat{G}$. There are natural counterparts of the canonical symplectic form $\Gamma$ on $V \times V^{*}$. It is clear that the most natural two-character on $\mathcal{G}=G\times \widehat{G}$, the analogue of the symplectic form of $V \times V^{*}$ ($V$ is a linear space) is given by 
\begin{equation}\label{eq.+10}
\zeta\left((x_{1}, \pi_{1}), (x_{2}, \pi_{2})\right)=\langle \pi_{1}| x_{2}\rangle\overline{\langle \pi_{2}| x_{1}\rangle}=\frac{\langle \pi_{1}| x_{2}\rangle}{\langle \pi_{2}| x_{1}\rangle}.
\end{equation}
This is a complex function on $\mathcal{G} \times \mathcal{G}$. This two-character is essentially non-singular, because, the mappings
\begin{equation}\label{eq.+11}
\mathcal{G}\ni (x, \pi) \mapsto \zeta((x, \pi), \cdot))\in \mathcal{G}, \qquad \mathcal{G}\ni (x, \pi) \mapsto \zeta( \cdot, (x, \pi)))\in \mathcal{G}
\end{equation}
are isomorphisms of $\mathcal{G}$ onto $\mathcal{G}$. Obviously, more rigorously, they are isomorphisms of $\mathcal{G}$ onto $\widehat{\mathcal{G}}$, but it is clear that $\widehat{\mathcal{G}}$, the dual of $\mathcal{G}$ is canonically isomorphic with  $\mathcal{G}$ itself, just like the linear spaces $(V \times V^{*})^{*}$ and $V \times V^{*}$ are canonically isomorphic.

In analogy to (\ref{eq.76a}) it might seem that we should have used $\zeta^{2}$ like in $V \times V^{*}$. However, in general, on the group $G$ such $\zeta^{2}$ fails to be non-singular. And it fails just in the case of groups $G$ like $\mathbb{Z}^{n}$ or ${\rm U}(1)^{n}$, physically interesting ones.

Wave functions are defined on $G$ in the "position representation"
and on $\widehat{G}$ in the "momentum representation". In analogy
to (\ref{eq.26}), (\ref{eq.27}), (\ref{eq.29}) for any $x\in G$ and any $\pi\in\widehat{G}$ we define linear operators of translation, $U(x)$, $V(\pi)$:
\begin{equation}
\left(U\left(x\right)\psi\right)\left(y\right)=\psi\left(x^{-1}y\right),
\qquad\left(V\left(\pi\right)\psi\right)\left(y\right)=\left\langle \pi|y\right\rangle \psi\left(y\right).\label{eq.+12}
\end{equation}
Obviously, 
\begin{equation}
\left(V\left(\pi\right)\psi\right)^{\wedge}\left(\lambda\right)=
\widehat{\psi}\left(\pi^{-1}\lambda\right).\label{eq.+13}
\end{equation}
The basic commutation relation is as follows \cite{Woron}:
\begin{equation}
U(x)V(\pi)U(x)^{-1}V(\pi)^{-1}=\overline{\left\langle \pi|x\right\rangle }=\left\langle \pi|x\right\rangle ^{-1}.\label{eq.+14}
\end{equation}
We can try to follow the line from $G=\mathbb{R}^{n}$ (additive)
and $\mathcal{G}=G\times \widehat{G}=\mathbb{R}^{2n}$ (additive):
\begin{equation}
W_{p}(x,\pi)=\left\langle \pi|x\right\rangle ^{p}U(x)V(\pi)=\left\langle \pi|x\right\rangle ^{p-1}V(\pi)U(x).\label{eq.+15}
\end{equation}
If in $G$ or $\widehat{G}$ does exist the unique square-rooting
operation (it does not exist in $\mathbb{Z}^{n}$ or ${\rm U}(1)^{n}$),
then putting $p=1/2$ we obtain the nice rule:
\begin{equation}
W\left(x^{-1},\pi^{-1}\right)=W\left(x,\pi\right)^{-1}.
\end{equation}
However, for topological groups different from $\mathbb{R}^{n}$,
$\mathbb{R}^{2n}$ something like this does not exist.

Let us consider the linear shell of operators $W_{p}(x,\pi)$; obviously,
the unitarity is meant in the sense of $L^{2}(G)$.

So in analogy to $G=\mathbb{R}$, we construct the family of operators:
\begin{equation}
\mathbf{A}:= \int \widehat{A}(x, \pi) \mathbf{W}_{p}(x, \pi)dx d\pi, 
\label{eq.+16}
\end{equation}
where $\widehat{A}$ is the Fourier transform of $A$; the both functions are defined  on the group $G \times \widehat{G}$, canonically isomorphic with $\left(G \times \widehat{G}\right)^{\wedge}$. 

For the corresponding composition of functions $A$, $B$ on the "phase space" $\mathcal{G} = G \times \widehat{G}$ we have the rule (\ref{eq.+1}), (\ref{eq.+4}) with the kernel function:
\begin{eqnarray}
&&K_{p}\left((x_{1}, \pi_{1}), (x_{2}, \pi_{2})\right)= \nonumber\\
&& \int \left\langle \pi_{1} | \xi \right\rangle \left\langle \eta | x_{1} \right\rangle \left\langle \pi_{2} | \zeta \right\rangle \left\langle \theta | x_{2} \right\rangle \left\langle \eta | \zeta \right\rangle^{1-p} \left\langle \theta | \xi \right\rangle^{-p} d\xi d\eta d\zeta d\theta. \label{eq.+17}
\end{eqnarray}

One can ask what would be the corresponding "$H^{+}$-canonical basis", corresponding to (\ref{eq.125a}), (\ref{eq.125b}) in the special case of $G= \mathbb{R}^{n}$, $\mathcal{G} = G \times \widehat{G} = \mathbb{R}^{2n}$. It would have to satisfy
\begin{equation}
A \star \rho_{x,y}= A(x) \rho_{x,y}, \qquad \rho_{x,y} \star A= \rho_{x,y} A(y) \label{eq.+18}
\end{equation}
for any function $A$ independent of "canonical" momenta as elements of $\widehat{G}$. 

If we define it in such a way, we obtain finally that
\begin{equation}
\rho_{x_{1},x_{2}}(x, \pi)= \delta\left(x_{1} x_{2} x^{-2}\right)\left\langle \pi | x_{1} x^{-1} \right\rangle. \label{eq.+19}
\end{equation}
This is strange in a sense. If $(x_{1} x_{2})$ is not the square of something, and it need not be so in a general $G$, then the above expression does vanish and is not a basis element. At this stage we are unable to answer the question what would be the basis of the corresponding "superselection" rule (\ref{eq.+19}), i.e., whether it is the bad choice of the procedure of finding this "canonical base" or just a real superselection for the quantum dynamics based on the von-Neumann equation:
\begin{equation}\label{eq.+20}
\frac{\partial \rho}{\partial t}= \frac{1}{i\hbar}\left(H \star \rho - \rho \star H \right).
\end{equation}

\section*{Acknowledgements}

This paper partially contains results obtained within the framework of the research project 501 018 32/1992 financed from the Scientific Research Support Fund in 2007-2010. The authors are greatly indebted to the Ministry of Science and Higher Education for this financial support. The support within the framework of Institute internal programme 203 is also greatly acknowledged.


\begin{thebibliography}{99}

\bibitem{AbrMar}
R. Abraham, J. Marsden, {\it The Foundations of Mechanics}, Benjamin Press, 1967, Addison-Wesley, 1978.
\bibitem{Arnold}
V.I. Arnold, {\it Mathematical methods of classical mechanics}, Springer Graduate Texts in Mathematics, Vol. 60,
Springer-Verlag, New York – Heidelberg, 1978.
\bibitem{Bargmann}
V. Bargmann. Ann. Math. 59, pp. 1–26, 1954.
\bibitem{LL}
L. D. Landau, E. M. Lifshitz, {\it Course of Theoretical Physics. Vol. III. Quantum Mechanics}, Pergamon Press, London, 1958.
\bibitem{Loomis}
L. H. Loomis, {\it An Introduction to Abstract Harmonic Analysis}, D. Van Nostrand Company, Inc., Princeton-New Jersey-Toronto-London-New York, 1953.
\bibitem{Mack}
G. W. Mackey, {\it The Mathematical Foundations of Quantum Mechanics}, Benjamin, New York, 1963.
\bibitem{Moyal}
J.E. Moyal, {\it Quantum mechanics as a statistical theory}, Proceedings of the Cambridge Philosophical Society, 45, 99–124 (1949).
\bibitem{Rubinowicz}
A. Rubinowicz {\it Quantum mechanics} Amsterdam, New York [etc.] Elsevier Pub. Co.; Warsaw, PWN-Polish Scientific Publishers, 1968 [1969]
\bibitem{Rudin}
W. Rudin, {\it Fourier Analysis on Groups}, Interscience Publ., New York-London, 1962.
\bibitem{Synge1}
J. L. Synge, {\it Classical Dynamics} in Handbuch der Physik, edited by S. Fl\..{u}gge, Springer-Verlag, Berlin, 1960. 
\bibitem{Synge2}
J. L. Synge, {\it Geometrical Mechanics and de Broglie Waves}, Cambridge Univ. Press, London, 1954.
\bibitem{JJS-phase}
J. J. S\l awianowski, {\it Geometry of Phase Spaces}, PWN --- Polish Scientific Publishers, Warsaw; John Wiley \& Sons, Chichester-New York-Brisbane-Toronto-Singapore, 1991.
\bibitem{JJS-Heis}
J. J. S\l awianowski, {\it Uncertainty, Correspondence and Quasiclassical Compatibility}, in: "The Uncertainty Principle and Foundations of Quantum Mechanics" A Fifty Years Survey. A tribute to Professor Werner Heisenberg to commemorate the fiftieth anniversary of the formulation of quantum mechanics, editors: W.C.Price, S.S.Chissick
John Wiley \& Sons, London-New York-Sydney-Toronto, pp. 147-188, 1977.
\bibitem{Weyl}
H.  Weyl, {\it The Theory of Groups and Quantum Mechanics}, Dover, New York, 1950.
\bibitem{Weyl2}
H. Weyl, {\it Quantenmechanik und Gruppentheorie },  Zeitschrift f\''{u}r Physik, 46, pp. 1-46  1927.
\bibitem{Wigner}
E.P. Wigner, {\it On the quantum correction for thermodynamic equilibrium}, Phys. Rev., 40, pp. 749–-759, 1932.
\bibitem{Woron}
S. Woronowicz, {\it On a theorem of Mackey, Stone and v.Neumann} Studia Mathematica 24, pp. 101-105 (1964).

\end{thebibliography}
\end{document}